\documentclass[sigconf,balance=false]{acmart}
\usepackage{hyperref}
\usepackage{popets}
\usepackage{graphicx}
\usepackage{array} 
\usepackage{booktabs}
\usepackage{hyperxmp}
\setcopyright{popets}
\copyrightyear{2025}
\acmYear{YYYY}
\acmVolume{YYYY}
\acmNumber{X}
\acmDOI{XXXXXXX.XXXXXXX}
\acmISBN{}
\acmConference{Proceedings on Privacy Enhancing Technologies}
\begin{document}

\title[Web Privacy based on Contextual Integrity]{Web Privacy based on Contextual Integrity: Measuring the Collapse of Online Contexts}


\author{Ido Sivan-Sevilla}
\orcid{0000-0003-2194-5006}
\affiliation{%
  \institution{The University of Maryland}
  \city{College Park}
  \state{Maryland}
  \country{USA}}
\email{sevilla@umd.edu}

\author{Parthav Poudel}
\affiliation{%
  \institution{The University of Maryland}
  \city{College Park}
  \state{Maryland}
  \country{USA}}
\email{ppoudel1@umd.edu}



\begin{abstract}
The collapse of social contexts has been amplified by digital infrastructures but surprisingly received insufficient attention from Web privacy scholars. Users are persistently identified within and across distinct Web contexts, in varying degrees, through and by different websites and trackers, losing the ability to maintain a fragmented identity. To systematically evaluate this structural privacy harm, we operationalize the theory of Privacy as Contextual Integrity and measure persistent user identification within and between distinct Web contexts. We crawl the top-700 popular websites across the contexts of health, finance, news \& media, LGBTQ, eCommerce, adult, and education websites, for 27 days, and created network graphs to learn how persistent browser identification via third-party cookies and JavaScript fingerprinting is diffused within and between Web contexts. Past work measured Web tracking in bulk, highlighting the volume of trackers and tracking techniques. These measurements miss a crucial privacy implication of Web tracking - the collapse of online contexts. Our findings reveal how persistent browser identification varies between and within contexts, diffusing user IDs to different distances, contrasting known tracking distributions across websites, and conducted as a joint or separate effort via cookie IDs and JS fingerprinting. Our network analysis informs the construction of browsers' storage containers to protect users against real-time context collapse. This is a first modest step in measuring Web privacy as Contextual Integrity, opening new avenues for contextual Web privacy research.
\end{abstract}

\maketitle
\section{Introduction}
Digital infrastructures have increased the scope and frequency of the collapse of social contexts \cite{10.1093/oso/9780198833659.003.0018}. It is now almost impossible for data subjects to maintain a fragmented identity online. This is especially evident on the Web, where ubiquitous tracking has been an inevitable part of the Web's structure, fueling the 225 billion dollar online advertising industry \cite{IAB.2024}. Surprisingly, even though the intentional collapse of multiple contexts into one has been so pervasive for Web users, it has not received sufficient scholarly attention. We lack a systematic measure of Web context collapse.

Past empirical work studied Web tracking in bulk, indirectly addressing context collapse. Scholars provided an essential window into the widespread uses of tracking practices, but had not provided a nuanced comparative understanding of context collapse. Scholars highlighted the prevalence of certain trackers and the centralized nature of the advertising industry. Comparative findings on tracking across websites are sparse, mainly revealing invasive tracking by news websites along with lower amounts of tracking on public websites, based on different incentives to monetize content.

We argue, in contrast, that users’ engagement with websites should not be regarded as one, all-encompassing, ’information-seeking’ context. Users' interactions with health, LGBTQ, adult, or educational websites for instance, should not be regarded as one context, but as distinct contexts in which the user wears separate identities. The collapse of these contexts into one is a fundamental breach of users' privacy and there are no notice and consent mechanisms that are effective enough in explaining and allowing users to protect themselves, leaving individuals exposed to pervasive context collapse when engaging on the Web.

To fill this gap, we operationalize the theory of Privacy as Contextual Integrity (CI). According to CI, profiling users based on information gleaned across social contexts, fails to respect contextual informational norms, and in so doing, violates people’ privacy expectations. Accordingly, we examine how persistent browser identification by stateful and stateless mechanisms is happening by third-party trackers, both within and between different online contexts. We assess when and by whom the norms of the original context are being violated and context collapse is taking place.

We collected stateful and stateless browser identification patterns across top 700 popular websites for 27 days, in seven different online contexts – Health, Finance, Education, LGBTQ, Adult, News \& Media, e-Commerce. We offer a close empirical examination, based on client-side interactions with websites, of the extent that trackers use persistent identifiers for users across and within contexts. We assume, based on previous works \cite{Olejnik2012WhyJC,browsinghistories2021}, that once a user ID is linked by trackers across different contexts, browsing history can be used to infer sensitive attributes about a person – health condition, desire to quit a job, political orientation, and etc. Even though we cannot observe how advertisers use data to decide on the best targeting method per user, we know that there is a natural incentive in this ecosystem to aggregate data in order to learn large fraction of user’s history and potential future behavior for targeting purposes \cite{255662, 10.1145/2660267.2660347, McGuigan.2019}.

We show which online contexts are more vulnerable to single and multiple context(s) collapse. We detail how the amount of trackers and participating websites varies between multiple and single context collapse, how the distribution of trackers across websites differs from the known, long tail distribution of trackers on the Web, and how far user IDs travel between online contexts. We also reveal the extent of overlap between the usage of cookie IDs and JS fingerprinting for context collapse. We then visualize the collapse of contexts via network graphs and calculate the associated chromatic numbers to inform the construction of browser's cookie storage containers and prevent real-time context collapse for users.

Specifically, the paper makes the following contributions:
\begin{itemize}
            \item \textbf{Develops a novel Web privacy measurement}: We designed a robust method to identify persistent browser identification and the collapse of single and multiple contexts on the Web, offering a novel measurement of Web privacy through the operationalization of the theory of privacy as Contextual Integrity, which rarely gets attention from Web privacy scholars.

            \item \textbf{Reveals the structural features of context collapse on the Web}: We show, for the first time, how persistent browser identification varies within and between Web contexts, highlighting the structural features that determine the level of real-time context collapse on the Web.

            \item \textbf{Informs browsers' storage containers}: We continuously monitor persistent identification patterns of the top popular websites across contexts and use network analysis techniques to inform the development of browsers’ containers that split cookies' storage to prevent context collapse.
\end{itemize}

In the next section we motivate our research by explaining our argument on how online context collapse takes place on the Web and why related work has yet to systematically evaluate and assess this privacy violation. We then turn to operationalize Web privacy based on CI, discuss our measurement framework, data collection, analysis, limitations, results, contribution, and suggestions. We conclude with future research avenues and open questions on contextual measurement of Web privacy.

\section{The Collapse of Online Contexts on the Web}

Digital infrastructures have been amplifying the collapse of social contexts \cite{10.1093/oso/9780198833659.003.0018}. Infrastructures such as social media platforms and the Web itself constantly blur and merge social contexts into one, lacking spatial, social, and temporal boundaries, making it difficult to maintain distinct social contexts when engaging online \cite{boyd2012}. 

We follow Davis and Jurgenson (2014) to define context in terms of role identities and their related networks. Context, then, refers to the identity meanings activated through interaction within a particular network of actors. Users are made up of multiple identities, each of which exists within a network of others and hold certain expectations that inform appropriate – and inappropriate – lines of action and identity performance. In these terms, context collapse refers to the overlapping of role identities through the intermingling of distinct networks \cite{doi:10.1080/1369118X.2014.888458}.

The collapse of online contexts has been mostly studied by social media and platform researchers, who demonstrated how these infrastructures blur the public and the private, make content available beyond the temporal moment of its creation, and obscure the viewership and unlimited audience for online content \cite{boyd2012,doi:10.1080/08838151.2012.732140,Wesch.2009}. Surprisingly, context collapse on the Web and its associated privacy implications has not received sufficient scholarly attention and yet to be measured systematically.

The intentional collapse of multiple contexts into one audience regularly occurs on the Web. Unlike our offline interactions, where we barely take information out of its original context and would not imagine, for example, reading private messages over the shoulder of a stranger \cite{Berjon.2021}, our online interactions, and specifically our interactions with websites, are full of intentional context collapse. Users’ Web history and site behavior seeps beyond the target website itself to external trackers for marketing purposes. Such context collision is accessible to marketers, often against users’ expectations \cite{Pew.2023}, happening both within- and between- different Web contexts, and enabling trackers to stitch together sensitive information about users’ queries and site interests from different web contexts. The majority of users are ‘not comfortable’ with such tracking and profiling, and do not want their data to be used for purposes other than proving the service they requested \cite{10.1145/1866919.1866929}. They are especially uncomfortable with actors navigating their data between different contexts to get a better understanding of what users might be interested in \cite{inmoment2018}. Notice and consent mechanisms fail to properly protect users \cite{Barocas2014BigDE,doi:10.1509/jppm.14.139}, leaving them exposed to constant collapse of their distinct online contexts. 

Empirical Web privacy researchers have been studying trackers in bulk, inspecting interactions between users' browsers and thousands of websites, with little sensitivity to the different contexts for users on the Web or to degrees of context collapse by the actors involved. Millions of third-party requests were examined by researchers to highlight the centralized nature of the advertising ecosystem and the significant presence of ‘top-trackers’ (e.g. Google, Meta, X) across the majority of measured traffic. The distribution of third-parties across the Web was found to have a long tail, with the majority of trackers operate on less than 1 percent of the Web \cite{Karaj2018WhoTracksMeMT,Yang2020,10.1145/2976749.2978313,solomos2020clashtrackersmeasuringevolution,Lerner2016,libert2015exposinghiddenwebanalysis,LiberBinns2019,Blacklight.2020,browsinghistories2021}. Users' Web history collected by third-party trackers was found to be unique to users and useful for extracting insights about users \cite {Olejnik2012WhyJC}. Researchers showed how users can be re-identified via their Web history, with the dominant trackers - Google \& Meta - unsurprisingly enjoying a privileged position to do so \cite{browsinghistories2021}.

Scholars who did consider different categories of websites showed variance in the amount of tracking within categories. An overarching finding was that news websites enable more third-party tracking than other types of websites, as opposed to public websites that demonstrate less presence of trackers \cite{10.1145/2976749.2978313, LiberBinns2019, 10.1145/3366423.3380203,Karaj2018WhoTracksMeMT, Yang2020, Lerner2016}. The difference in the amount of tracking for each category is usually associated with the incentives for publishers to include third-party trackers. Different business models and funding resources across websites are hypothesized as a possible explanation for the observed variance of tracking amount across website categories \cite{10.1145/2976749.2978313}.

Hu et al. (2020) conducted research that is the closest we could find to a direct assessment of online contexts collapse on the Web \cite{Hu2020}. They examined how first-party websites are connected based on the common third-parties they use. They found that selectively removing the largest trackers is a very effective way of decreasing the interconnectedness of websites. Another important, albeit indirect assessment of context collapse was conducted by Sorensen and Kosta (2019), who examined the shift in third-party typology on the Web for the course of eight months, before and after the GDPR \cite{BeforeandAfterGDPR.2019}. They found similarities with earlier works showing the dominance of giant tech companies as third-parties and a long tail distribution of third parties across the Web, highlighting how public websites have far fewer third-parties than private ones.

Empirically, we diverge from existing research by pivoting our observations of website interconnectedness based on distinct contexts, and spot patterns of persistent identification that go beyond common presence of a third-party among first-party websites, since not all third-party trackers are participating in persistent user identification.

Existing studies do not distinguish persistent browser identification patterns within and between Web contexts and the extent to which they enable the collapse of online contexts. In contrast, our approach enables to engage with questions such as Which third-parties contribute to the collapse of online contexts? Who is responsible for context collapse beyond the 'usual suspects' of advertising giants? What are the websites that are more or less vulnerable to online context collapse? We lack a comparative understanding across popular web contexts of how using persistent user identifiers differs among contexts. For example, do users' interactions with popular healthcare websites are more/less commonly linked to users’ browsing behavior in other popular Web contexts? To what extent the fact that users browse to consume adult content can serve advertisers when the same users engage in e-commerce, seek financial advice, or look for health advice? We aim to conduct a context-sensitive analysis of persistent browser identification across the Web and gain a better understanding of how and by whom users' contexts are collapsing.

Alarmingly, context collapse has become a pervasive feature of our online lives, shaping what is known about us and forms our epistemic practices, habits, and opportunities \cite{10.1093/oso/9780198833659.003.0018}. The tendency to collapse online contexts risks a main feature enabled by privacy - the ability of individuals to hold multiple identities in different social contexts. Identity is by nature fragmented: we present ourselves as different people in different contexts \cite{Berjon.2021}. The collapse of online contexts, however, leads to overlapping individual identities through the intermingling of distinct networks.

To measure the collapse of online contexts on the Web, and systematically understand how the third-party structure of the Web has been enabling these privacy harms, we turn to the theory of privacy as Contextual Integrity (CI) and develop a novel measurement of Web privacy.
\subsection{Web Privacy as Contextual Integrity}

Nissenbaum (2004) argues that our expectations of privacy are contextual. Different social contexts are characterized by different norms of what information it is appropriate to reveal within that context and how information should be transferred from one party to another \cite{Nissenbaum2004}. Privacy is violated, according to CI, when the norms of the context are violated and observed information flows do not advance the purpose for the context to exist in the first place. In the Web example, user queries in healthcare websites, LGBTQ websites, or eCommerce websites are differently tagged to a context, and hold different purposes for the user. We search healthcare websites when we seek a health advice and improve our health condition. We tour LGBTQ websites to learn and engage with communities based on gender interests. And we shop online on e-commerce websites to acquire products and services we would like to consume. 

Data, therefore, is never just data. Depending on the context, data can be healthcare data, educational data, or financial data, and each type of data should advance the purpose of its context. Healthcare data exist to improve health outcomes. Financial data flows should facilitate efficient allocation of capital. Gender-related data should promote gender interest and knowledge of data subjects. Each of these contexts has distinct guiding values and purposes.

Based on the category of websites, we can assess their context and evaluate the purpose of user engagement, tagging users’ web history and site interactions to a context. It is a violation of privacy to share information outside original web contexts in ways that do not advance the purpose for which these contexts exist for users. Notice and consent choices are unable to capture the ways users' Web history and site interaction data lead to the collapse of online contexts because the complex data flows for content monetization on the Web are beyond individual cognitive capacity \cite{Barocas2014BigDE,doi:10.1509/jppm.14.139}.

In times, violating contextual norms is virtuous since it advances a greater socially desired goal. Cases of whistleblowing, or the sharing of health information to mitigate a global pandemic can all be desired collapses of online contexts. But the collapse of contexts for user targeting and marketing purposes goes against users’ expectations and does not respect the norms in the original Web context. According to past surveys, users and consumers expect their online information flows to maintain the integrity of the context in which they operate \cite{Pew.2023}. 

Contextual Integrity is used to flag violation of contextual norms in two distinct browsing scenarios: (1) \textbf{within a web context}, when, for example users seek health advice across an array of healthcare websites, and persistent user identification by third-party trackers enables the construction of rich user profiles based on users’ Web history across sites from the same context. These users' site histories are revealing a great deal of users’ health concerns, and collecting them by an external third-party tracker for marketing purposes is collapsing the online context for users interacting with the array of healthcare sites. We refer to this as 'single context collapse.' (2) \textbf{Between contexts}, persistent browser identification further changes the meaning of sensitive information. It enables trackers to bring together users’ web history and interests from different online contexts, collapsing not just one, but two or more online contexts for the sake of marketing and advertising. For example, when users’ browsers are persistently identified during subsequent visits to healthcare and LGBTQ websites, users experience context collapse in both, allowing trackers and marketers to build user profiles based on user engagement in both contexts, without advancing the original purpose of these contexts. We refer to this as 'multiple contexts collapse.'

We seek to reach a better understanding of how and by whom online contexts collapse on the Web. Notably, CI does not flag specific types of data for special protection without specifying relevant contextual factors. It disagrees with privacy approaches that circumscribe particular categories of information for restrictive or laxer treatment without consideration of contextual factors. Contextual factors are characterized by five parameters of an information flow: data subject, sender, and recipient (collectively referred to as the actors), information type (or attribute), and transmission principles (the conditions that constrain data flow from senders to recipients).

We use CI's contextual factors to map and label the information flows we wish to assess, as detailed in Figure 1. In that figure, the interaction of the user with the first context is on the left, followed by the interaction of the user with another context on the right. Specifically, to measure multiple contexts collapse we inspect information flows in which: (1) the sender of the information is the user’s browser; (2) the recipient of the information is tracker A in \textbf{context Y}; (3) the information type is the assigned user ID by tracker A in \textbf{context X}; (4) the transmission principle is notice \& consent; and (5) the data subject is the individual who interacts with the website through the browser.

To measure within-context collapse we inspect information flows in which: (1) the sender of the information is the user’s browser; (2) the recipient of the information is tracker A in \textbf{Website 2} within context X; (3) the information type is the assigned user ID by tracker A in \textbf{Website 1} within context X; (4) the transmission principle is notice \& consent; and (5) the data subject is the individual who interacts with the website through the browser.

\begin{figure}[htp]
    \centering
    \includegraphics[width=1\linewidth]{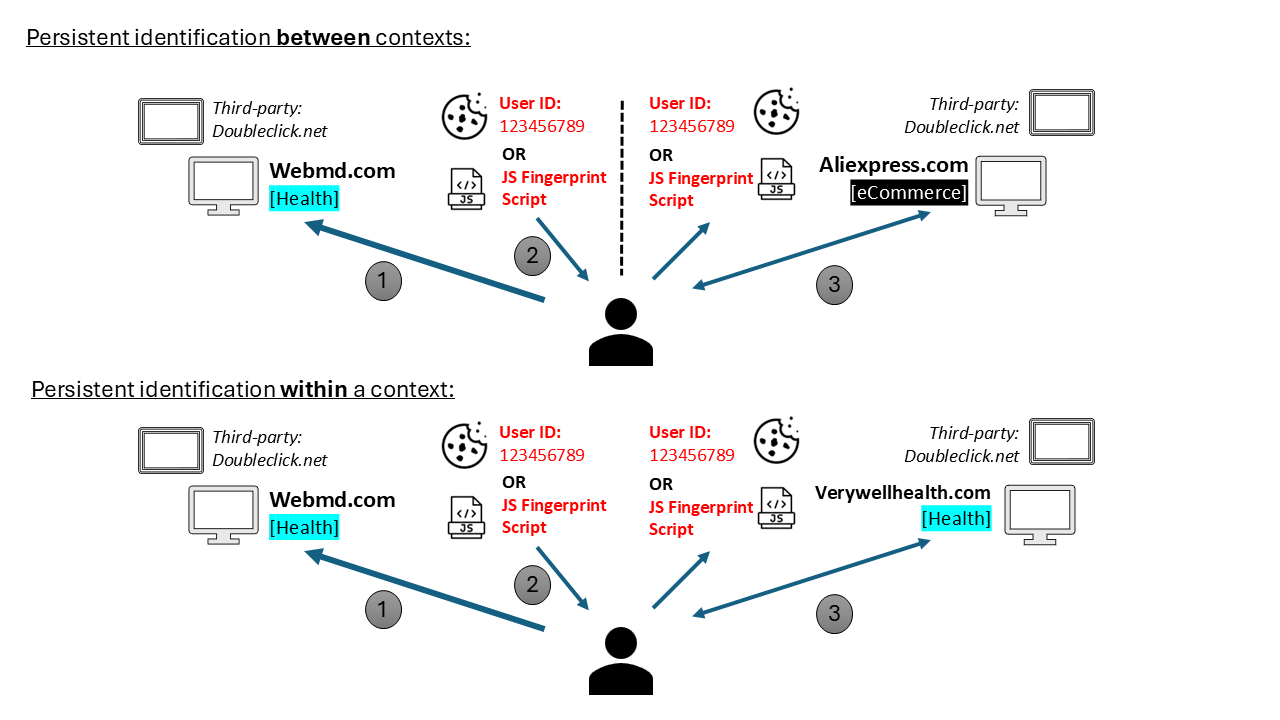}
    \caption{Information flows that capture multiple and single context(s) collapse}
    \label{fig:enter-label}
\end{figure}
We argue that these information flows represent a collapse of online contexts within or across Web Contexts. These flows provide the opportunity for tracker A to conflate contexts Y \& X, or for tracker A to conflate context X. To make this more concrete, we do not want our interactions with healthcare websites to be coupled with our interactions with LGBTQ or ecommerce websites for example, by actors who cannot support our health condition and for purposes that are beyond medical advice.

\section{Measurement Framework}
    Our measurement framework is based on spotting persistent browser identification within and across online contexts. We use instances of persistent browser identification by trackers as a proxy to the collapse of online contexts, either within or between different contexts. We then carefully pick website categories that constitute distinct contexts and crawl the top popular 700 websites across seven distinct categories (100 websites per category) to identify and analyze the trackers across sites that enable context collapse through the sharing of cookie IDs or recurring JS fingerprinting scripts within or between contexts.
    
    \subsection{Persistent User Identification on the Web}

    There are two distinct approaches to generate unique identifiers for users on the Web. One is ‘stateful,’ where the client’s browser saves identities locally, as a long string, usually via HTTP cookies or JavaScript APIs, and later retrieve these unique IDs to identify the same users across websites. The second type of ID generation is ‘stateless,’ and is based on information about the browser and/or network to create a unique ‘snapshot’ fingerprint of the user in a given moment based on browser’s type, canvas/font, web traffic, audio settings, and battery levels. These identifiers are not saved locally by clients’ browsers but are observed and probably saved by trackers \cite{10.1145/2872427.2882991,10.1145/2976749.2978313,Yang2020,Karaj2018WhoTracksMeMT}. 

    We pivot our empirical observations around both third-party identification cookies and JavaScript fingerprinting functions that assign a dynamic ‘advertising identifier’ to users as they browse the Web. Those ‘advertising Identifiers’ are a key prerequisite for the industry to address a user for personalized/targeted advertising since all user-centric data is associated with it \cite{IAB2020}. Despite the default block of third-party cookies by several browsers, Google, that operates the browser with the largest market share \cite{Statista.2024}, has recently announced that third-party cookies are here to stay, making the collapse of online contexts via cookie IDs still very relevant \cite{TomsGuide.2024}.
    
    \subsection{Case Selection}
        We test persistent browser identification trends across the top 100 popular websites in seven distinct online contexts (700 websites overall) - News \& Media, Health, Education, Finance, LGBTQ, eCommerce, and Adult. Each website category is constructed based on different purposes and values, carry different weights and identities for users, and can be uniquely tagged to a distinct context: User interactions with news \& media websites are based on the willingness to consume updated information or knowledge. Healthcare websites provide users with medical advice and serve as a space for users to advance their health. Educational websites signal users' interest and educational aspirations. Users' interactions with financial websites facilitate knowledge on efficient allocation of capital. Gender-related websites promote users' gender interests and knowledge. eCommerce websites enable users to acquire products and services they would like to consume, and adult websites signal users' sexual desires. These distinct contexts should not be exposed to trackers (=collapse within context) and should not be linked to other contexts from the same user (=collapse between contexts).

        Following our selection of categories, we selected the top 100 popular websites from each of the seven categories (700 websites overall) based on the \textit{similarweb.com} list of popular websites. This provided a comprehensive cross-section list of heavily visited sites across different contexts. Table 1 below indicates the average number of visitors from the top 100 websites in each context category. The Appendix section provides the full list of websites crawled for each context. We chose websites based on their popularity to learn about context collapse in the commercial Web, where we already know that tracking patterns pervasively take place.
        \begin{table}[htp]
            \begin{tabular}{lc} \toprule
            \textbf{Website Context}  & \textbf{Average Monthly Visits} \\ \midrule
                 News \& Media& 46,128,960 \\ 
                 Health& 13,716,310 \\ 
                 Education& 3,246,934 \\ 
                 Finance& 18,016,110 \\ 
                 LGBTQ& 219,651 \\ 
                 eCommerce& 44,260,900 \\ 
                 Adult& 47,391,330 \\ \bottomrule
            \end{tabular}
            \caption{Average unique monthly visits across top-100 popular websites in each context (from \textit{similarweb.com})}
            \label{tab:my_label}
        \end{table}
\section{Data collection} 
   
    Through an open-sourced instrumented Firefox browser, OpenWPM, that simulates browser activity and records website responses, metadata, cookies used, and scripts executed \cite{10.1145/2976749.2978313}, we studied the top 700 popular websites across seven Web contexts. We ran seven crawls per day for the 700 websites, starting with a different context every time, to learn how user identification is further diffused by trackers from the first context to other contexts. Each context was a priority context and crawled first, followed by the other categories in a random order. We ran seven crawls of the selected 700 websites per day, for 28 days, From Oct 26 - November 28 of 2024, from an AWS server located at Ashburn, Virginia, USA. 

    We studied stateful and stateless tracking practices via HTTP \& JavaScript cookies and JavaScript APIs. For each experiment, we are matching cookie IDs among contexts to realize which trackers, and from which websites, user IDs are diffused across contexts. As opposed to previous studies, we do not assume that the presence of trackers in two different contexts means that they persistently identify users across those contexts \cite{10.1145/3394231.3397897}. Instead, we looked for a valid evidence, i.e. the usage of the same cookie-id across contexts, to assume persistent user identification. This leaves us only with trackers who engaged in uniquely identifying users' browsers.

    To improve the accuracy of our data collection efforts and ensure that the crawled websites are revealing their 'true' tracking behavior to our bot crawler, we implemented several recommended modifications to OpenWPM from Krumnow et al. (2022) \cite{KJK22}:

    \begin{itemize}
        \item \textbf{Stealth JavaScript Instrumentation}: By default, many websites can detect OpenWPM and similar tools, which might result in them adjusting their behavior. In response, we enabled a stealth mode for the JavaScript instrumentation, allowing our automated browser to be less detectable. 
        \item \textbf{Navigator Webdriver Override}: Selenium, the underlying framework of OpenWPM, sets the \texttt{navigator.webdriver} property to \texttt{true}, making the tool identifiable as an automated browser. We changed this value to \texttt{false} to further mask the bot-like characteristics of the crawling tool.
        \item \textbf{Native Display Mode}: Instead of using the more common "headless" browser mode, we opted for a "native" display mode with a full graphical user interface. This approach makes the browsing session appear more like a real user and decreases the likelihood of detection.
        \item \textbf{Dynamic Window Resolutions}: We randomized window and screen resolutions across different browsing sessions, emulating the diverse range of devices used by actual users. This technique added another layer of trust to the browsing behavior of our bot.
        \item \textbf{Bot Detection Mitigation}: Natively, OpenWPM includes a bot mitigation feature designed to mimic human interaction with websites (random scrolling, clicking, and resizing of the page). This feature prevents sites from blocking our crawlers based on automated behaviors.
    \end{itemize}

    To allow for complete page load and the setting of all cookies, we configured each crawl to "sleep" for 10 seconds after visiting a page. This sleep time ensured that all page resources were fully loaded before moving to crawl the next website.

    \subsection{Collecting Patterns of Persistent Browser Identification and Context Collapse}

        During the seven crawling experiments per day, we collected both stateful and stateless browser identification patterns. For stateful browser identification via third-party cookies, we collected two types of cookies: (1) HTTP Cookies - traditional cookies set by servers through HTTP headers using the \texttt{Set-Cookie} directive. These cookies are typically stored in the browser and sent with every subsequent HTTP request to the same domain; (2) JavaScript Cookies - these cookies are set or manipulated directly by a JavaScript on the client side, typically via the \texttt{document.cookie} API. Unlike HTTP cookies, JS cookies can be fully controlled by the website’s JavaScript, enabling dynamic interactions with the user session. Based on previous studies, we identified an ID cookie by the following four elements \cite{Yang2020,10.1145/2660267.2660347,10.1145/2976749.2978313}:
        \begin{itemize}
            \item \textbf{Cookie lifetime}: The cookie has a lifetime longer than 90 days.
            \item \textbf{Parameter-Value length}: The length of the parameter-value string is greater than 7 but less than 101.
            \item \textbf {Parameter-Value remains similar within a crawl}.
            \item \textbf {Parameter-Value varies between crawls}: Utilizing the Ratcliff-Obershelp algorithm, we selected cookies for which the parameter-value is less than 66\% similar among experiments conducted on the same website in different days.
        \end{itemize}
        Each cookie that was recognized based on these requirements was then manually checked in the tracker's privacy policy to validate that it is indeed an ID cookie used by this tracker to uniquely identify the user's browser. These steps ensured that we have one ID cookie per each tracker observed in our data.

        To collect stateless browser identification patterns, we used existing work \cite{Fingerprinting2021} to spot JavaScript API keywords that are likely to be related to fingerprinting scripts. A tracker that applies the same JavaScript fingerprinting scripts in two websites can link user records from both, persistently identifying the user's browser within or across contexts. To recognize JS fingerprinting scripts, We followed the criteria on Iqbal et al. (2021), according to which frequently used JavaScript API keywords in fingerprinting scripts have to appear in three websites and should be at least 16 times more likely to be used in fingerprinting rather than non-fingerprinting scripts \cite{Fingerprinting2021}. Importantly, despite being a stateless browser identifier, applying fingerprinting scripts on subsequent browsing experiences is likely to produce the same unique fingerprint from the same combination of browser and device configurations.

        Both stateful and stateless browser identifiers were collected from the first crawled context, and then examined across subsequent crawls in other contexts. The crawl of 700 websites, seven times a day, was conducted over the period of 28 days to test the validity of our findings and check whether differences in the quantity of tracking between contexts are statistically significant. The data collection and analysis processes are visualized in Figure 2.

    \begin{figure}[htp]
        \centering
        \includegraphics[width=1\linewidth]{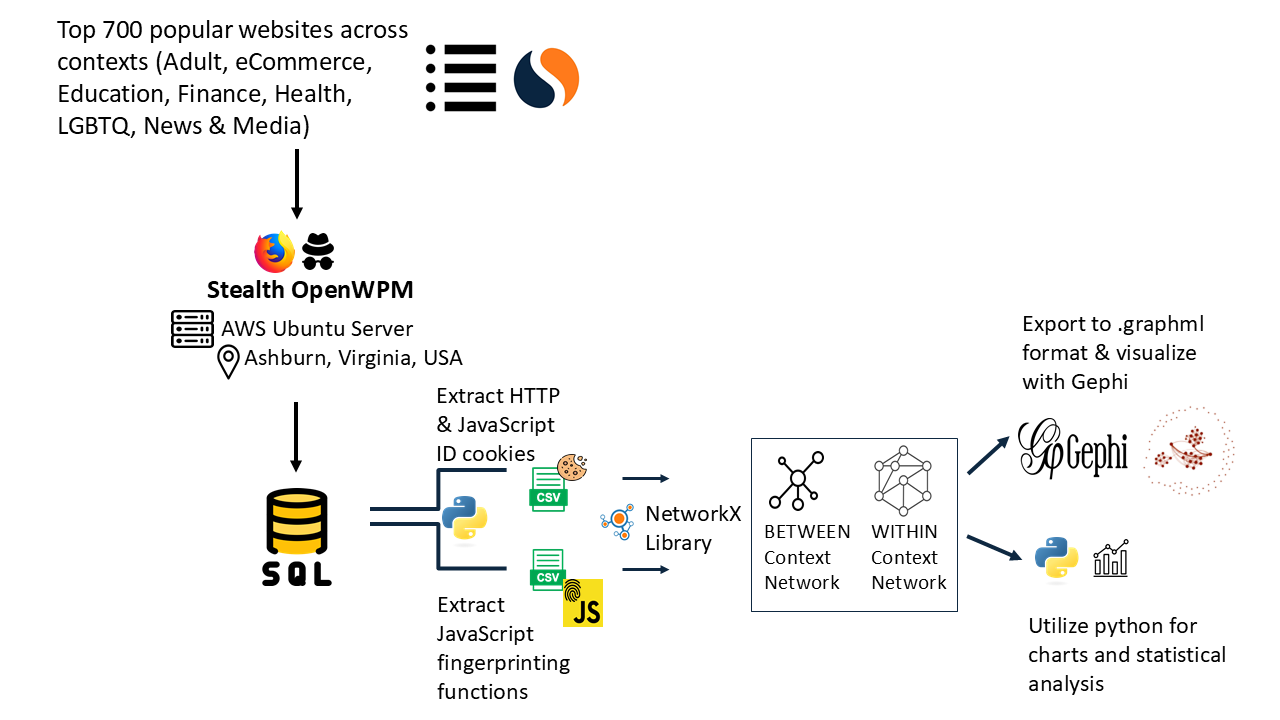}
        \caption{Data Collection Workflow}
        \label{fig:enter-label}
    \end{figure}

\section{Limitations}
    Due to various limitations, our results should be considered as a lower bound for persistent browser identification patterns and context collapse within a single or across multiple contexts on the Web. The following list details our limitations and areas for future engagement.
    \begin{itemize}
            \item \textbf{Crawling only homepages}: We only crawled the homepages of selected websites to avoid bot detection. It is known that most tracking is happening on inner rather than homepages of websites \cite{10.1145/2976749.2978313,10.1145/3366423.3380203}, but at the same time, previous work found that crawling inner pages increases the chances of OpenWPM discovery \cite{KJK22}.
            
            \item \textbf{Studying clear-text ID cookie values}: We are aware that cookie values can be encoded or encrypted when used by the same tracker in different websites \cite{Fouad2020}. For our study, we considered only identical ID cookie values observed during our crawls. We could not decrypt or find in our data cookie IDs that could be easily decoded.

            \item \textbf{Not fully considering cookie-syncing practices}: In our study of stateful user identifiers, we conducted a preliminary analysis of how trackers share user IDs via cookie syncing and forwarding practices, including through the use of tracking pixels. ID sharing between trackers is another information flow that leads to context collapse. Even if different user IDs are assigned for different websites, in different contexts, the linkage of IDs can be observed via HTTP requests in the browsing session \cite{Fouad2020,Papadopoulos_2019}. 
            Our preliminary analysis reveals that sharing of user IDs varies significantly between contexts, across both the number of trackers and participating websites. We spot interesting relations between trackers who already identified the user's browser and trackers that did not directly drop an ID cookie on our browser. We plan to conduct an in-depth follow-up study in a separate paper about the impact cookie syncing and forwarding have on context collapse.
            
            \item \textbf{Single geolocation for the crawling}: We repeatedly crawled our selected websites from a single geolocation in Virginia, USA. Previous studies showed how crawling from different geolocations might produce different results \cite{BeforeandAfterGDPR.2019}. Our results, therefore, mimic the browsing experience from North America and should be only regarded as such.

            \item \textbf{Crawling top-popular websites}: We chose to crawl top-popular websites, with tens of millions of user visits per website every month (see Table 1), to learn how context collapse happens in the popular Web. At the same time, we are not measuring context collapse based on users' real browsing behavior. This requires a dedicated browser extension and agreement from users to share their data. We aim to develop this in future research efforts.
        \end{itemize}

\section{Results}
We examined the interconnectedness of a context within and across other contexts. For each context, we crawl the context's top-100 popular websites first, before moving to other sets of contexts in the same crawl. We examine how and by whom browser identification travels from the first context to other contexts (=multiple context collapse), as well as how websites are connected to one another within the same context (=single context collapse). For each context we had 28 different observations from 28 days of crawl. The tables below present the average values across all days of crawling. The visualized instances of context collapse are based on one crawl that had the closest-to-average number of participating websites in persistent browser identification across all days.

\subsection{Multiple context collapse: Persistent user identification between contexts}

Table 2 presents average statistics of multiple context collapse across different context origin. The first row, for example, describes the crawl of top 700 popular websites when the top-100 popular adult websites were crawled first. The first row indicates that on average, 285.76 unique third-parties appeared in the adult context, and an average number of 461.8 cookies were dropped on the browser. 56.8 third-parties (19.88 percent of all third-parties in the adult context) were labeled as 'persistent identifiers' because they persistently identified the browser from the Adult context to one or more of the following contexts in that crawl. 9.04 percent of all crawled adult websites, on average, enabled persistent browser identification from their websites to other contexts. Differences between the mean of persistent identifiers from each context and participating websites from each context were found to be statistically significant based on single-factor ANOVA tests as detailed in the appendix.

Unsurprisingly, and consistent with previous findings, the total number of third-parties and cookies vary between contexts, with News \& Media websites hold the highest number for both. We see different numbers of persistent identifiers and participating websites across contexts. 56.88 percent of popular News \& Media websites carry user IDs to other contexts, with only 9.04 percent of Adult websites and 39.01 percent of Health websites do so, on average, for their contexts. News \& Media websites embed the highest number of persistent identifiers in their websites and enable persistent identification through the majority of websites crawled. Financial websites are not very far behind, and then Health, eCommerce, and LGBTQ websites. Education is second to last, and Adult websites embed the lowest number of persistent identifiers and participating websites. 

Interestingly, eCommerce, health, and LGBTQ websites have similar numbers of persistent identifiers (130-147 on average). Health websites have close to 40 percent of websites that enable context collapse from their sites to subsequent contexts. The percentage of trackers that are 'persistent identifiers' between contexts is similar across contexts and range from 18.5 to 24.27 percent of the total third parties in that context. Similar portions of third parties in each context are labeled as persistent identifiers, showing surprising similarity in this tracking pattern across contexts.

\begin{table*}
    \centering
    \begin{tabular}{@{}>{\raggedright\arraybackslash}p{0.1\linewidth}>{\raggedright\arraybackslash}p{0.1\linewidth}>{\raggedright\arraybackslash}p{0.1\linewidth}>{\raggedright\arraybackslash}p{0.1\linewidth}>{\raggedright\arraybackslash}p{0.1\linewidth}>{\raggedright\arraybackslash}p{0.1\linewidth}>{\raggedright\arraybackslash}p{0.1\linewidth}@{}}\toprule
    First Context& Average unique third-parties& Average number of cookies& Average of Persistent Identifiers& Percentage of Persistent Identifiers & Participating Websites in Context Collapse\\
    \midrule
         Adult&  285.76&  461.8&  56.8&  19.88\%& 9.04\%\\ \hline 
         eCommerce&  597.2&  1276&  137.28&  22.99\%& 28.57\%\\ \hline 
         Education&  431.92&  681.44&  100.52&  23.27\%& 23.52\%\\ \hline 
         Finance&  767.88&  1969.96&  186.4&  24.27\%& 43.31\%\\ \hline 
         Health&  669.32&  1323.64&  147.08&  21.97\%& 39.01\%\\ \hline 
         LGBTQ&  601.64&  1077.36&  130.8&  21.74\%& 30.42\%\\ \hline
         News \& Media& 995.8& 2811.92& 184.44& 18.52\%& 56.88\%\\
    \bottomrule
    \end{tabular}
    \caption{Average statistics of between-context collapse based on ID cookies across 25 days of crawl}
    \label{tab:my_label}
\end{table*}

Table 3 describes how persistent browser identification travels from the first context to other contexts in the crawl. It shows how far user identifiers diffuse to subsequent contexts, providing an assessment of how persistent those persistent identifiers are. Interestingly, the distance that persistent identifiers travel from their original context varies between contexts. Originating from the Adult context, the majority of persistent identifiers (56.36 percent) travel, on average, all the way to the last context in the crawl. Education is next, for which 34.1 percent of persistent identifiers originated in the education context travel all the way. For Health and LGBTQ contexts, the highest portion of persistent identifiers stops after five contexts. In absolute numbers, however, 30-40 persistent identifiers from each and every context, on average, travel all the way from the original context to the last context in the crawl.

\begin{table*}
    \centering
    \begin{tabular}{@{}>{\raggedright\arraybackslash}p{0.07769999999999999\linewidth}>{\raggedright\arraybackslash}p{0.07769999999999999\linewidth}>{\raggedright\arraybackslash}p{0.07769999999999999\linewidth}>{\raggedright\arraybackslash}p{0.07769999999999999\linewidth}>{\raggedright\arraybackslash}p{0.07769999999999999\linewidth}>{\raggedright\arraybackslash}p{0.07769999999999999\linewidth}@{}>{\raggedright\arraybackslash}p{0.07769999999999999\linewidth}>{\raggedright\arraybackslash}p{0.07769999999999999\linewidth}}\toprule
    First Context& Average Number of Persistent Identifiers& One Context Diffusion
& Two Contexts Diffusion& Three Contexts Diffusion& Four Contexts Diffusion &Five Contexts Diffusion &Diffusion across all contexts\\
    \midrule
         Adult&  56.5&  3.88\%&  4.79\%&  7.37\%& 9.40\%& 18.20\%&56.36\%\\ \hline 
         eCommerce&  136&  7.67\%&  9.73\%&  12.42\%& 16.05\%& 25.30\%&28.83\%\\ \hline 
         Education&  100&  4.95\%&  7.32\%&  10.83\%& 13.93\%& 28.88\%&34.10\%\\ \hline 
         Finance&  186&  10.19\%&  13.55\%&  16.61\%& 15.23\%& 20.43\%&23.99\%\\ \hline 
         Health&  146&  6.76\%&  9.42\%&  14.81\%& 17.83\%& 26.02\%&25.15\%\\ \hline 
         LGBTQ&  130&  7.28\%&  10.66\%&  13.68\%& 16.17\%& 26.29\%&25.93\%\\ \hline
         News \& Media& 184& 12.85\%& 14.68\%& 16.23\%&16.30\%& 19.64\%&20.29\%\\
    \bottomrule
    \end{tabular}
    \caption{Distance traveled, on average, by Cookie IDs from the original to subsequent contexts}
    \label{tab:my_label}
\end{table*}

Figures 3 \& 4, from different crawls that started from a different context, illustrate this difference. In the visualized crawl in Figure 3, 60.49 percent of persistent identifiers that originated from the Adult context traveled all the way to the last context - LGBTQ. 18.52 percent of the identifiers stopped after six contexts (including the original one). The rest of the identifiers (less than 20 percent in total) diffused to three or less subsequent contexts. For persistent identifiers originated in LGBTQ websites, however, illustrated in Figure 4, the diffusion is more even, making the visualization more colorful. 20.78 percent traveled all the way to the websites in the Finance context. The majority of identifiers (27.7 percent) traveled through six contexts (including the origin), 21.05 percent of the identifiers traveled through five contexts (including the origin). The rest 30 percent of identifiers stopped after 3 contexts or less.

\begin{figure}[htp]
    \centering
    \includegraphics[width=1\linewidth]{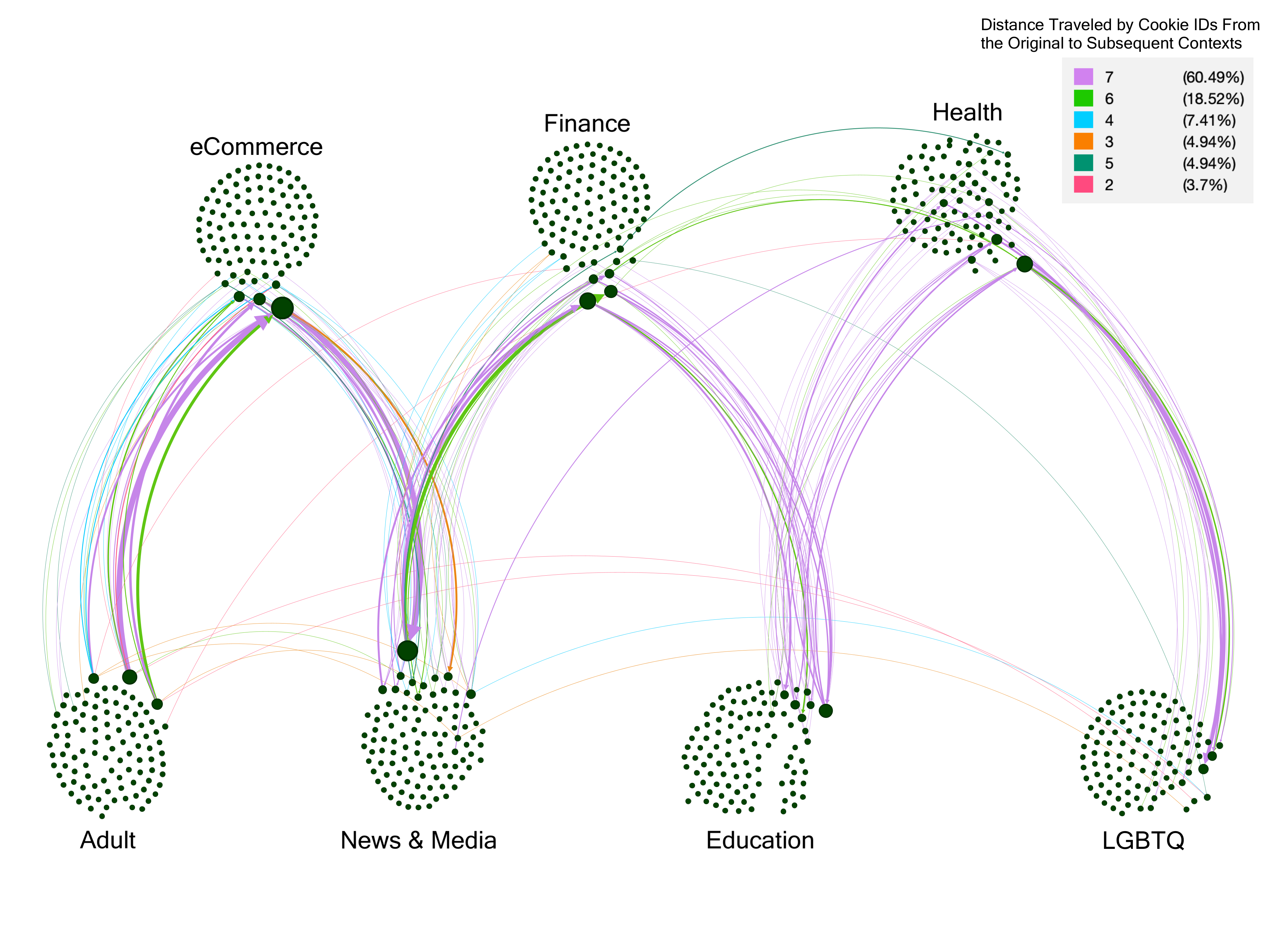}
    \caption{The Travel of Persistent Identifiers from the Adult Context on Oct 26, 2024}
    \label{fig:enter-label}
\end{figure}

\begin{figure}[htp]
    \centering
    \includegraphics[width=1\linewidth]{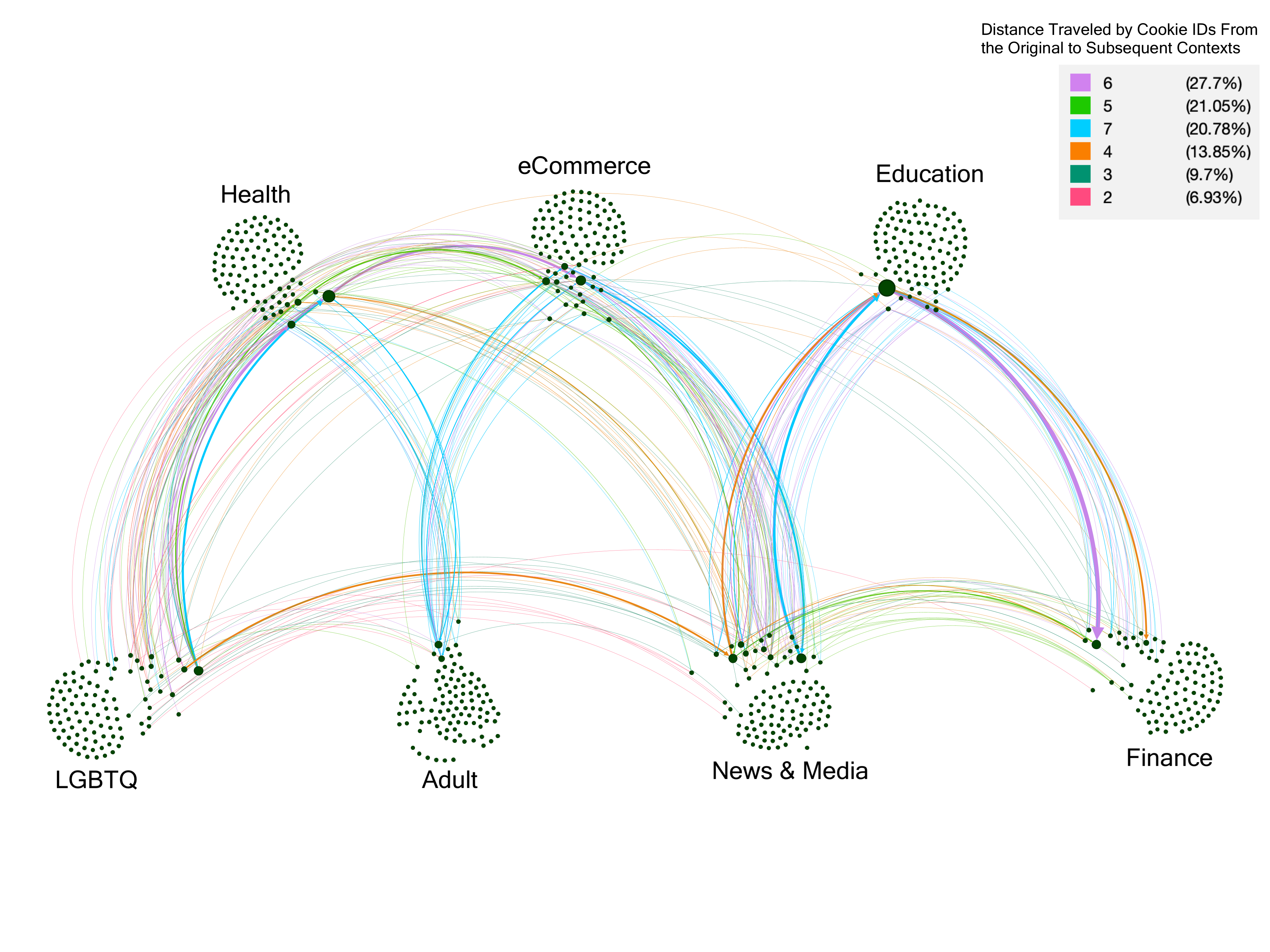}
    \caption{The Travel of Persistent Identifiers from the LGBTQ Context on Oct 26, 2024}
    \label{fig:enter-label}
\end{figure}

We now move to inspect the coverage of websites by trackers that act as persistent identifiers. Previous work found a long-tail distribution for third-parties across the Web, with the vast majority of third-parties cover less than one percent of the websites that were studied \cite{BeforeandAfterGDPR.2019,10.1145/2976749.2978313}. In contrast, we found that persistent identifiers coverage holds a short tail distribution across the websites they connect between contexts. Figure 5 shows the different distribution for each context, based on different context of origin. From six contexts, the highest percentage of websites covered by one tracker is 20 percent, while the rest of the trackers cover 0-10 percent of the websites across contexts. The adult context is an outlier, with a few trackers cover 30-45 percent of the connected websites from this context. This finding shows that persistent identification across multiple contexts does not follow known tracking trends that were found to be more centralized and dominated by a few trackers.

\begin{figure}[htp]
    \centering
    \includegraphics[width=1\linewidth]{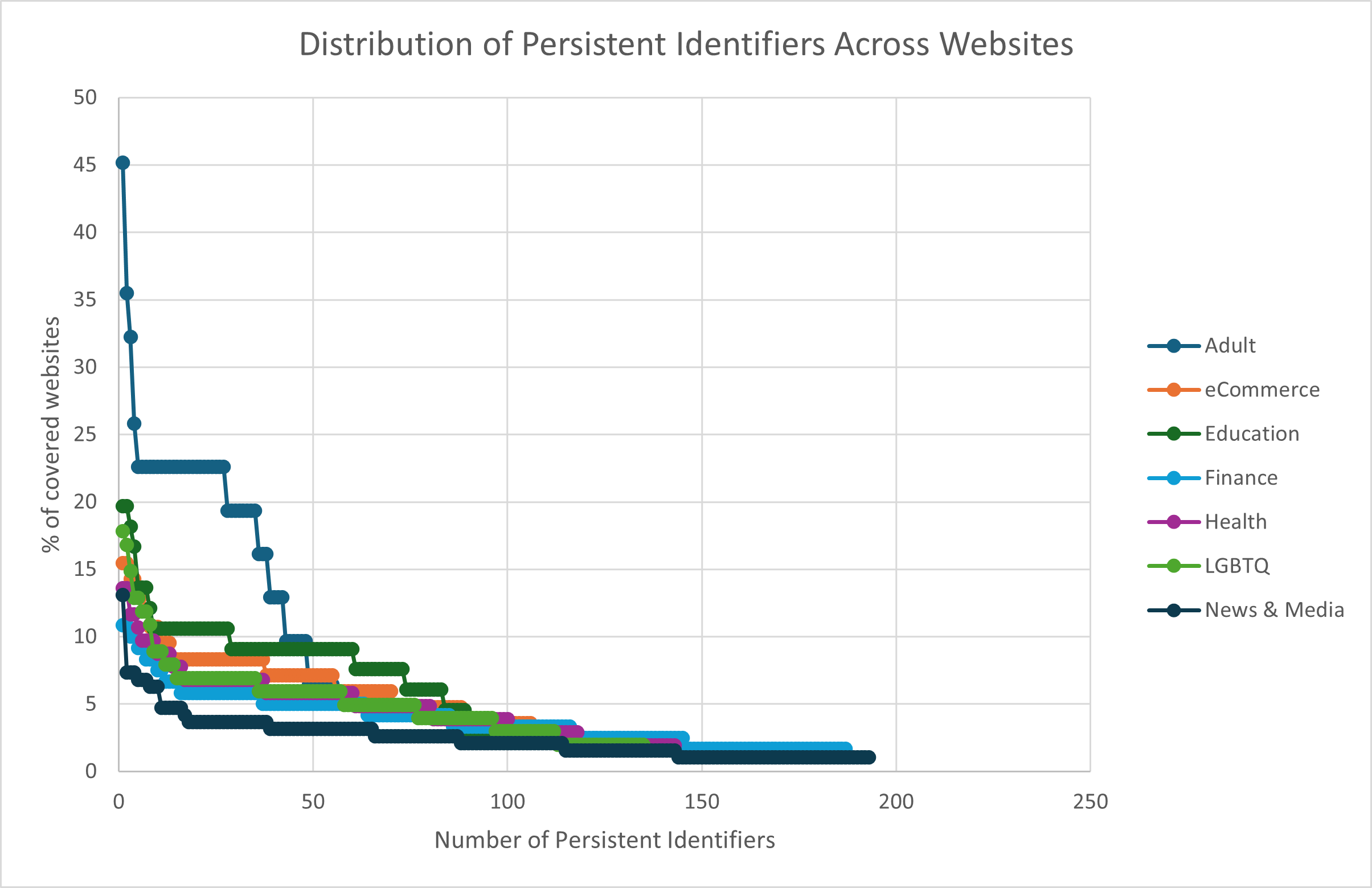}
    \caption{Distribution of Persistent Identifiers Between-contexts Across Websites}
    \label{fig:enter-label}
\end{figure}

For each context, there are a few dominant trackers and notable connected websites worth highlighting. Figure 6 visualizes multiple context collapse from the Finance context. Trackers that are labeled as persistent identifiers are in the middle, connecting colored nodes that represent websites. The size of the nodes change based on their in- and out-degree. The crawling took place clock-wise, with the 12 o'clock context crawled first. We can see how persistent identification takes place by many actors (187) and connecting more websites between contexts (120), relative to those numbers in other contexts. Specifically, (foxbusiness.com), (economictimes.com), (xe.com), and (toyokeizai.net) are the financial websites that mostly connect browser identification to other contexts. They are notably connected to (ratemyprofessors.com) in the education context, (ssg.com) and (slickdeals.net) from the eCommerce context, (businessinsider.com) and (oneindia.com) in the News \& Media context, (autostraddle.com) and (queerty.com) in the LGBTQ context, and (chemistwarehouse.com.au) in the health context. These sites are key participants in the collapse of online contexts for users originating from the Finance context. The dominant trackers, in the middle of the figure, that are linking websites from the finance context to other contexts are (bing.com) that links 13 websites, (taboola.com) that links 13 websites, (demdex.net) that links 12 websites, and (rubiconproject.com) that links 12 websites. As shown in the distribution of persistent identifiers across websites (Figure 5), the 187 persistent identifiers from the finance context are almost evenly distributed across the websites they connect.

\begin{figure}[htp]
    \centering
    \includegraphics[width=1\linewidth]{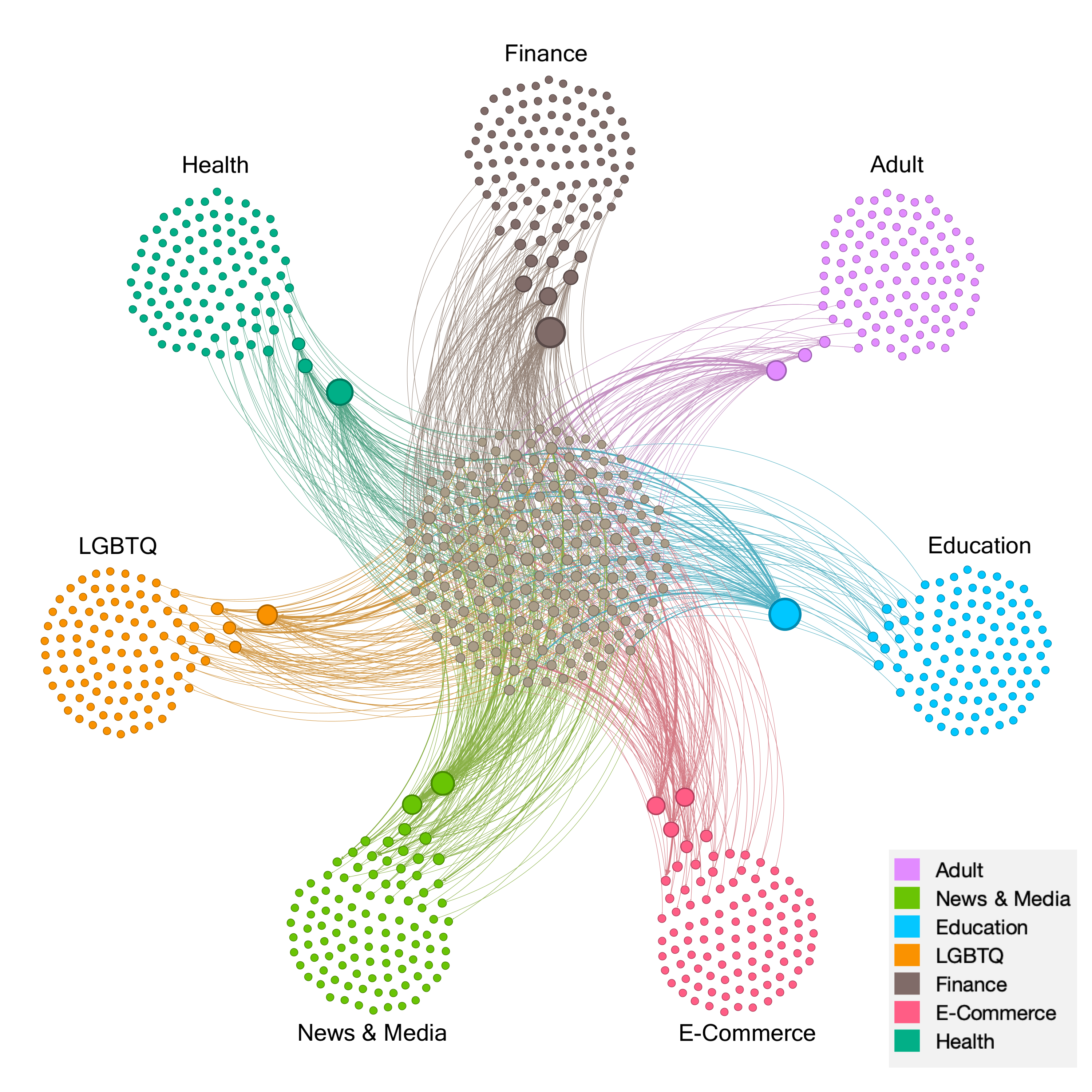}
    \caption{Trackers originate from the finance context connecting to websites from other contexts}
    \label{fig:enter-label}
\end{figure}

Surprisingly, a few spotted persistent identifiers serve as first-party websites in our top-700 popular websites list, significantly increasing their visibility on users. The list of those websites, their first-party web context, and their third-party tracking contexts are detailed in the appendix.

\subsection{Single Context Collapse: Persistent user identification within the context}

Table 4 presents average statistics of within-context collapse by trackers per context, over 25 days of collected data. The second row, for example, describes average results from all crawls of top 700 popular websites when the top-100 popular eCommerce websites were crawled first. 597 unique third-parties appeared in the eCommerce context on average, and an average number of 1276 cookies were dropped on the browser. 102 third-parties (17.1 percent of all third-parties in the eCommerce context) were labeled as 'persistent identifiers' because they persistently identified the browser across sites within the eCommerce context. 58.4 percent of all crawled eCommerce websites, on average, enabled persistent browser identification to other websites within that context.

As opposed to persistent identification between-contexts, we found that much less third-parties are persistently identifying the browser within the same context. The number of persistent identifiers within-context decreases by 20-50 percent for six contexts, and for News \& Media by less than 10 percent. The percentage of persistent identifiers within-context out of the total number of third-parties is lower as well, ranging from 10.55 percent in the adult context to 18.36 percent within the eCommerce context. The number of websites that participate in within-context collapse, however, has doubled, and sometimes even tripled, in comparison to between-context collapse. With 81.2 percent of News \& Media sites and 74.8 percent of finance sites on average are linked within their contexts. Differences between the mean for persistent identifiers within each context and for participating websites in each context are statistically significant, as shown in the single-factor ANOVA tests in the appendix.

\begin{table*}
    \centering
    \begin{tabular}{@{}>{\raggedright\arraybackslash}p{0.1\linewidth}>{\raggedright\arraybackslash}p{0.1\linewidth}>{\raggedright\arraybackslash}p{0.1\linewidth}>{\raggedright\arraybackslash}p{0.1\linewidth}>{\raggedright\arraybackslash}p{0.1\linewidth}>{\raggedright\arraybackslash}p{0.1\linewidth}@{}}\toprule
    Context& Average unique third-parties& Average number of cookies& Average of Persistent Identifiers& Percentage of Persistent Identifiers & Participating Websites in Context Collapse\\
    \midrule
         Adult&  285.76&  461.8&  30.16&  10.56\%& 32.44\%\\ \hline 
         eCommerce&  597.2&  1276&  102.36&  17.14\%& 58.4\%\\ \hline 
         Education&  431.92&  681.44&  55.36&  12.82\%& 67.48\%\\ \hline 
         Finance&  767.88&  1969.96&  140.96&  18.36\%& 74.84\%\\ \hline 
         Health&  669.32&  1323.64&  109.72&  16.4\%& 62.96\%\\ \hline 
         LGBTQ&  601.64&  1077.36&  102.48&  17\%& 57\%\\ \hline
         News \& Media& 995.8& 2811.92& 169.88& 17.1\%&81.2\%\\
    \bottomrule
    \end{tabular}
    \caption{Average statistics of within-context collapse based on ID cookies across 25 days of crawl}
    \label{tab:my_label}
\end{table*}

Figure 7 shows the distribution of persistent identifiers within contexts across the websites they connect. Results are different from what was found in the analysis of multiple context collapse. Within contexts, we found a long tail distribution of trackers - a few trackers connect most of the websites within the same context. The adult context, which had a longer tail distribution in the between-contexts analysis, is an outlier again, having a shorter tail distribution. The leading persistent identifiers within contexts are the 'usual suspects' from the advertising industry - (doubleclick.net) is the leading tracker within all contexts, with more or close to 50 percent visibility of websites within contexts. (adnxs.com), and (criteo.com) are also popular within the News \& Media and LGBTQ contexts. In the Health context we see (adsrvr.org) as another leading context collapse initiator. In Finance it is also (bing.com), in Education we also see (facebook.com) and (linkedin.com), and in Adult websites (yandex.ru) and (tsyndicate.com) persistently identify browsers across 10 and 7 percent of the websites.

\begin{figure}[htp]
    \centering
    \includegraphics[width=1\linewidth]{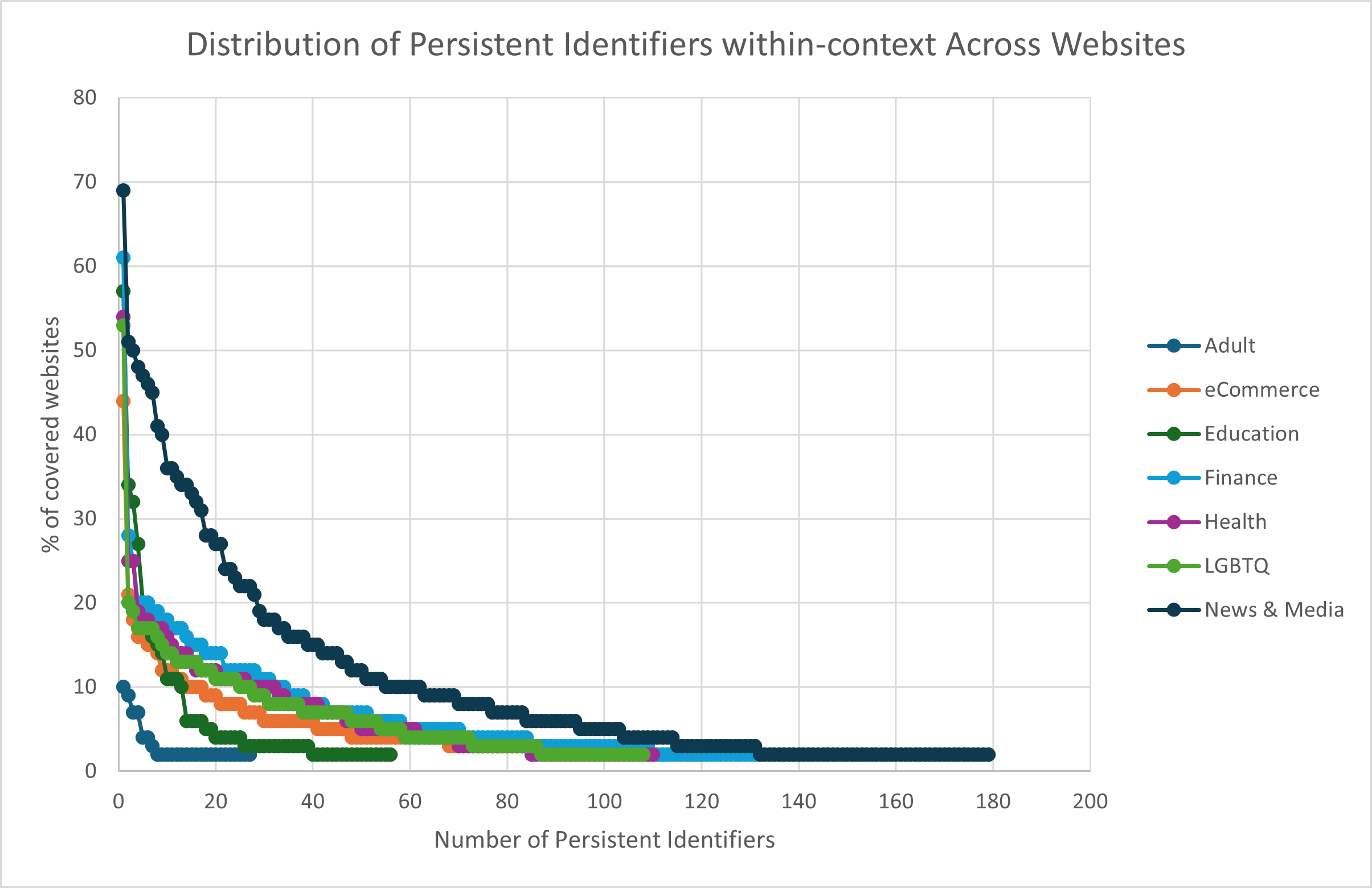}
    \caption{Distribution of Persistent Identifiers within-context Across Websites}
    \label{fig:enter-label}
\end{figure}

Figure 8 shows persistent browser identification within the education context. 71 percent of websites within this context have an out-degree in the network that is greater than one - meaning that the browser is persistently identified between those websites and other websites in that context. The leading websites with the highest out-degree are (asu.edu - 66), (colorado.edu - 64), and (tophat.com - 61). These websites are linking more than 60 percent of the websites crawled in this context. On the other end, the three websites with the highest in-degree are (gmu.edu - 64), (sydney.edu.au - 63), and (purdue.edu - 62). These are websites that get connected from more than 60 percent of the websites in the education context.

\begin{figure}[htp]
    \centering
    \includegraphics[width=1\linewidth]{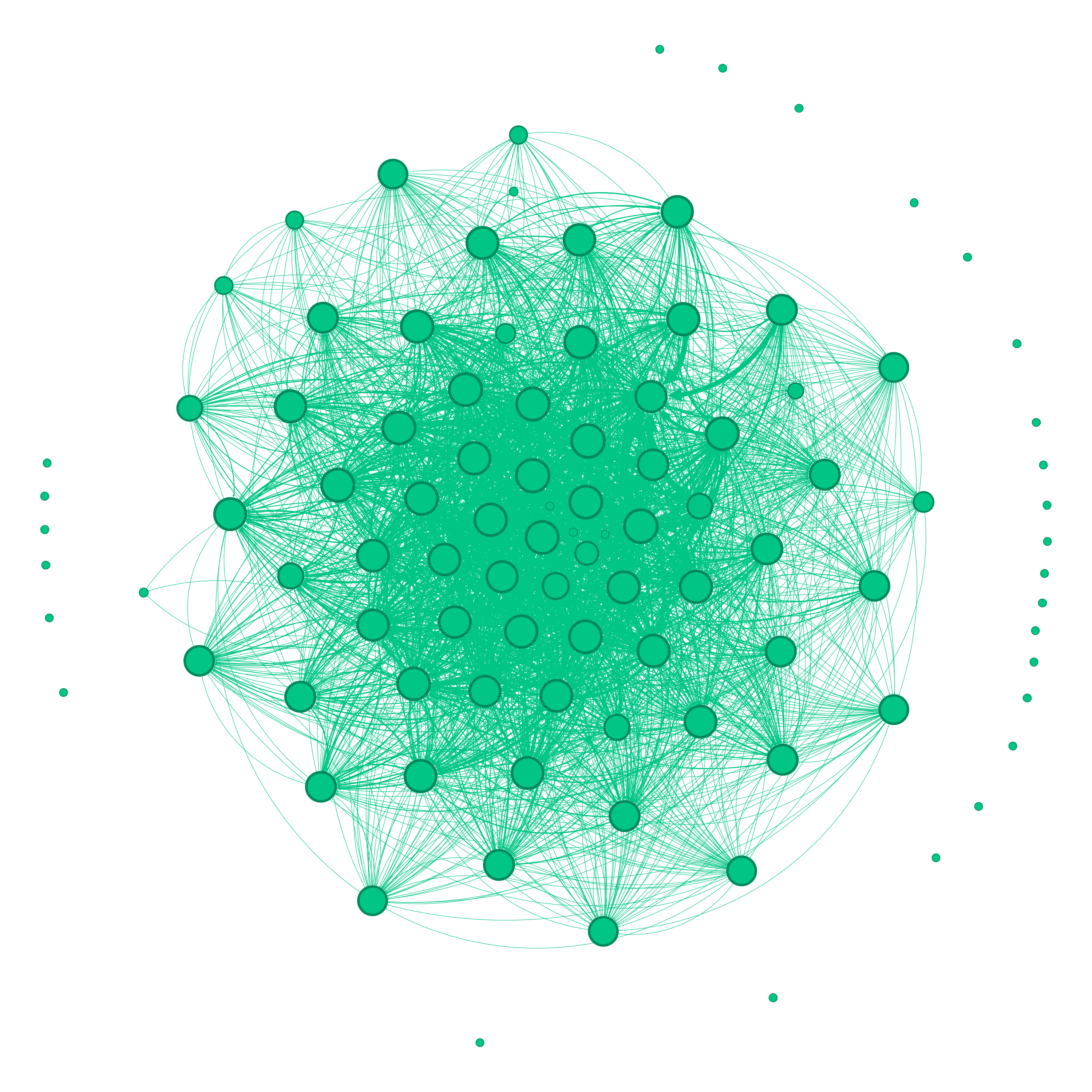}
    \caption{Within context connectivity in the education context}
    \label{fig:enter-label}
\end{figure}

The trackers that frequently connect websites within the education context are the leading, well-known trackers from the advertising industry. Figure 9 visualizes their work. The size of the nodes is determined by their degree. The four largest nodes are (doubleclick.net - connecting 57 sites), (facebook.com - connecting 34 sites), (linkedin.com - connecting 32 sites), and (youtube.com - connecting 27 sites).

\begin{figure}[htp]
    \centering
    \includegraphics[width=1\linewidth]{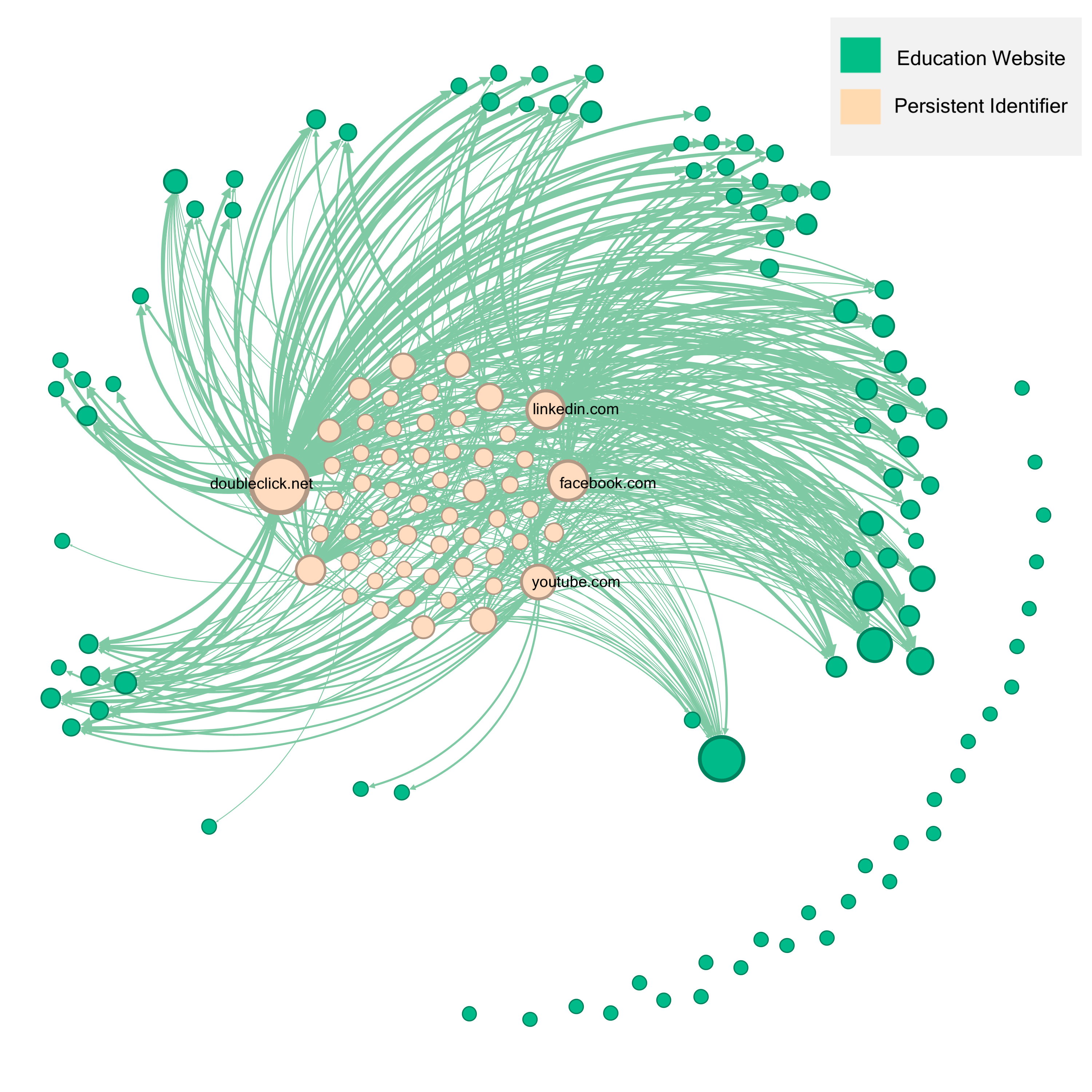}
    \caption{Persistent Identifiers connecting education websites}
    \label{fig:enter-label}
\end{figure}

\subsection{JS fingerprinting between- and within-contexts}
Next, we examined the use of JS fingerprinting by trackers to identify browsers between and within contexts. Given the increased user awareness and browsers' default actions against third-party cookies, we wanted to see to what extent JS fingerprinting is used by persistent identifiers as either 'backup plans' in case third-party cookies are not available or as a tracking instrument that breaks new grounds of context collapse. Those JS-based fingerprints remain consistent as long as users' system configuration stays the same, which can make them even more persistent than cookie IDs. We looked at crawls from specific dates to compare cookie ID and JS fingerprinting context collapse.

The usage of JS fingerprinting by trackers to persistently identify multiple or single context collapse was 5-10 times lower than the usage of cookie IDs, across all contexts under study. The number of participating websites connected via JS fingerprinting was 2-3 times lower than their number connected via cookie IDs. When inspecting overlap between cookie ID and JS fingerprinting usage by actors for multiple context collapse, we found 13 different trackers that use both cookie IDs and JS fingerprinting. They originated from six out of our seven studied contexts (none was found in the Adult context), and their applied JS fingerprinting scripts usually travel to fewer contexts than the cookie IDs they use. On average, 259 persistent identifiers use cookie IDs exclusively to collapse multiple contexts, and 24 persistent identifiers appear in our data for the first time as they exclusively use JS for multiple context collapse. Interestingly, 62 new websites can now be tagged as participating websites in context collapse using JS fingerprinting. These websites cover sites from all contexts, distributed almost evenly among contexts.

For single context collapse (within context analysis), we found 13 different trackers that use both tracking methods, with 11 of them overlapping with the JS trackers found in the between-context analysis. 250 persistent identifiers exclusively use cookie IDs for single context collapse, and 26 persistent identifiers appear for the first time as trackers that initiate within-context collapse. 24 new websites, evenly distributed across six contexts but News \& Media sites, are new participating websites in within-context collapse.

Figure 10 visualizes multiple context collapse from the eCommerce context on Nov 3, 2024. Grey nodes and edges represent context collapse that is exclusviely happening via cookie IDs. Orange nodes and edges represent user ID diffusion by both cookie IDs and JS fingerprinting. Nodes and edges in light blue are places where there has been an exclusive use of JS fingerprinting for multiple context collapse. Ten new edges and eighteen new nodes are exclusively connected via JS fingerprinting functions in this crawl. 

\begin{figure}[htp]
    \centering
    \includegraphics[width=1\linewidth]{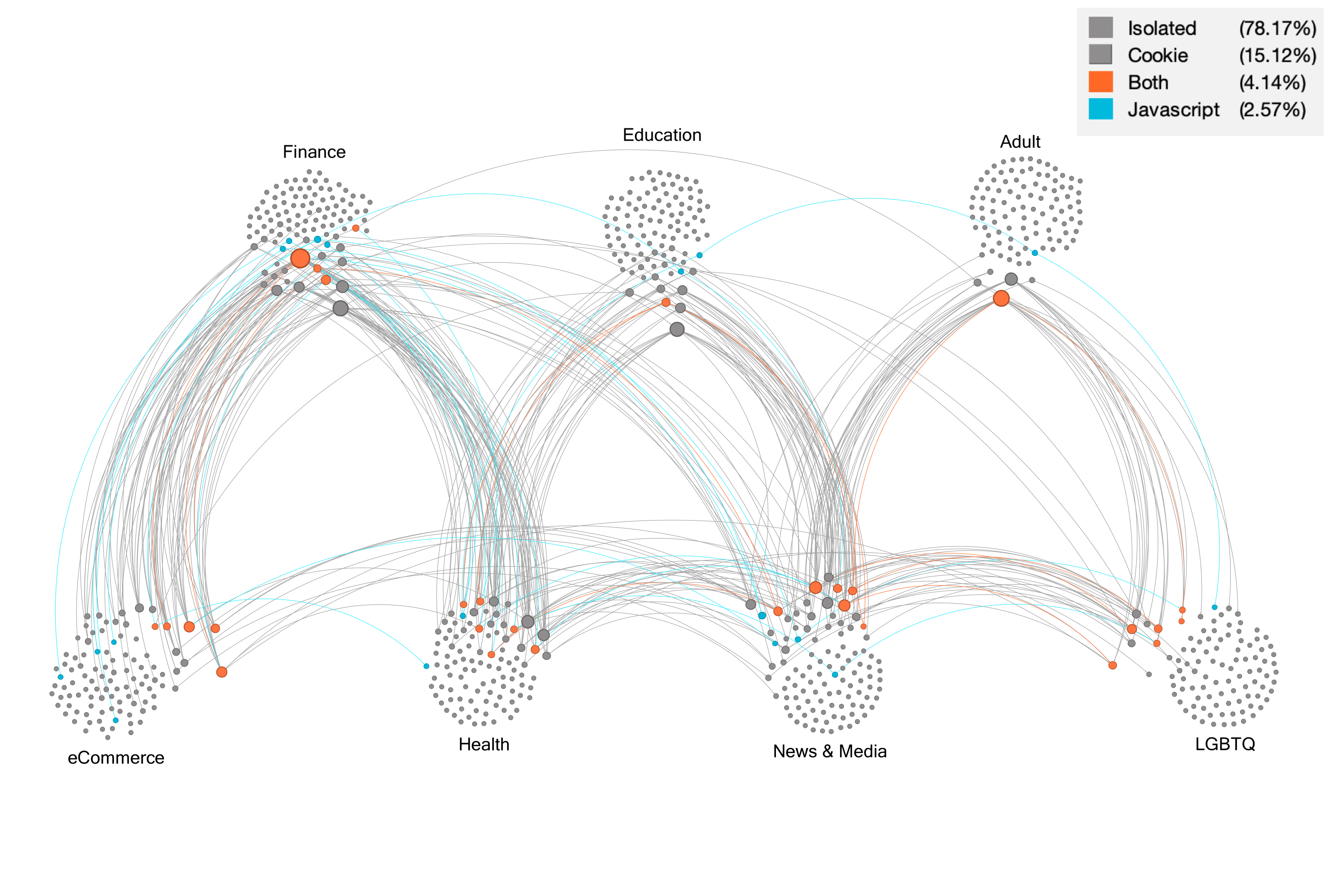}
    \caption{Multiple context collapse from the eCommerce context on Nov 3, 2024 by both cookie IDs and JS fingerprinting}
    \label{fig:enter-label}
\end{figure}

\section{Discussion \& Suggestions}

Our results reveal the varying structural connections of online contexts on the Web and show the extent of single and multiple context collapse, for tens of millions of users per month, on the top-700 popular websites across seven distinct contexts. The amount and portion of trackers that engage in persistent browser identification and the participating websites that embed those trackers vary between contexts. Single context collapse, the linkage of websites from the same context, tends to behave like traditional Web tracking, with many websites within a context are connected via dominant trackers from the advertising industry, and a few trackers cover most of the websites within each context. Multiple context collapse, however, is enabled by a fewer number of websites in each context, but by more trackers and on significant portions of enabling websites. From each context, the same percentage of third parties link user IDs further to other contexts, with most user identifiers travel to six and seven contexts.

Our between-contexts analysis shows the collapse of multiple contexts, highlighting variations based on the context of origin. The number of persistent identifiers and participating websites vary between the contexts, but at least 30-40 third-party trackers are persistent identifiers, and most of them diffuse user IDs to more than five other contexts. Our within-context analysis shows the collapse of a single context for users, that varies based on context of origin but mostly follows common tracking trends found before. The distribution of trackers across websites vary as well, with multiple context collapse suggests a short tail distribution, while single context collapse shows a long tail distribution for trackers across websites.

We found that the amount of trackers that use JS fingerprinting for real time persistent identification of browsers is 5-10 times lower than the number of persistent identifiers that use cookie IDs. Still, JS fingerprinting is used to boost and sometimes create exclusive connections between websites that were not connected via cookie IDs. The scale of that is low, but this is a context collapse trend worth following in the future.

Results show how prevalent single and multiple context collapse are, and how it changes based on context and type of context collapse on the popular Web. There are several hypotheses for why context collapse varies between contexts, linking the collapse to the Web personas of users \cite{8400224}, the persona of websites' owners \cite{tubiblio137333}, and the business models of websites \cite{BeforeandAfterGDPR.2019,10.1145/2976749.2978313}. We have not collected data to explain where various patterns of context collapse come from, as it goes beyond the scope of this paper and requires a mixed-methods approach. 

Our contextual measurement of Web privacy can be used to inform Mozilla's continuous efforts to advance contextual identities for users on the Web through separating browser's cookie and local storage to different containers \cite{Mozilla.2024}. Assigning first-party websites to different storage containers prevents trackers from accessing data associated with different contexts, practically preventing cookie ID-based persistent identification patterns found in this paper. Mozilla previously suggested a set of contexts for users based on their (1) personal; (2) work; (3) banking; and (4) shopping activities \cite{Mozilla.2022}. Instead, we call for a set of site-specific containers, per context, based on our dynamic analysis of persistent browser identification within- and between-contexts.

Per context, our results can be used to realize the required number of containers and the first-party websites to include in each container in order to prevent context collapse. We build on Hu and Sastry (2020) and further develop their 'Tangle Factor' to realize how first-party websites are connected, within and between distinct online contexts \cite{Hu2020}. We would like to open our data analysis outputs and inform the design of browsers' cookie storage containers by finding the vertex chromatic number and list of colored nodes of our between- and within-context network graphs. Per context, the number of colors needed for nodes such that neighboring nodes which share an edge are colored differently, provides the number of required containers. The list of nodes in each color provides the list of first-party websites to include in each container. Per context, the number of required containers and the list of first-party websites to include in each container can guide browser developers on how to break users' browsing experience into storage containers and prevent the collapse of that context for the user.

We demonstrate the applicability of our approach through the two figures. In Figure 11, we can see the connected websites within the LGBTQ context. The vertex chromatic number of the graph is 53, indicating that we need 53 different containers to ensure that users are not being persistently identified within the top-100 popular LGBTQ websites. The list of first-party websites to include under each container is provided in the appendix.

\begin{figure}[htp]
    \centering
    \includegraphics[width=1\linewidth]{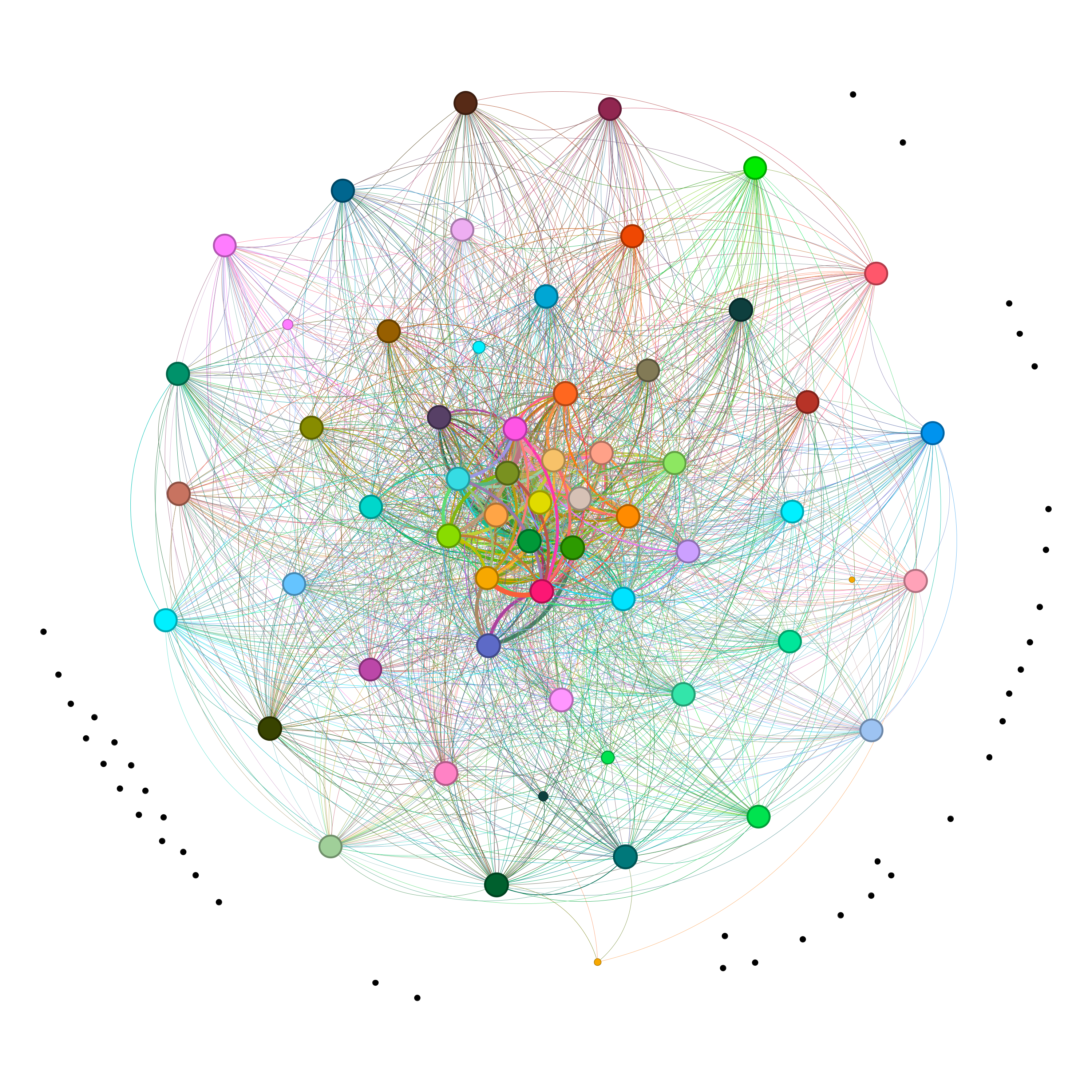}
    \caption{Within-context connections of LGBTQ websites, each vertex is colored based on the calculated chromatic number of the graph.}
    \label{fig:enter-label}
\end{figure}

In Figure 12, we can see the connected websites from multiple contexts, based on user identifiers created in the LGBTQ context. The graph shows that in order to prevent multiple contexts collapse, we need to have eight containers and include first-party websites according to their color, as shown in the appendix. 

\begin{figure}[htp]
    \centering
    \includegraphics[width=1\linewidth]{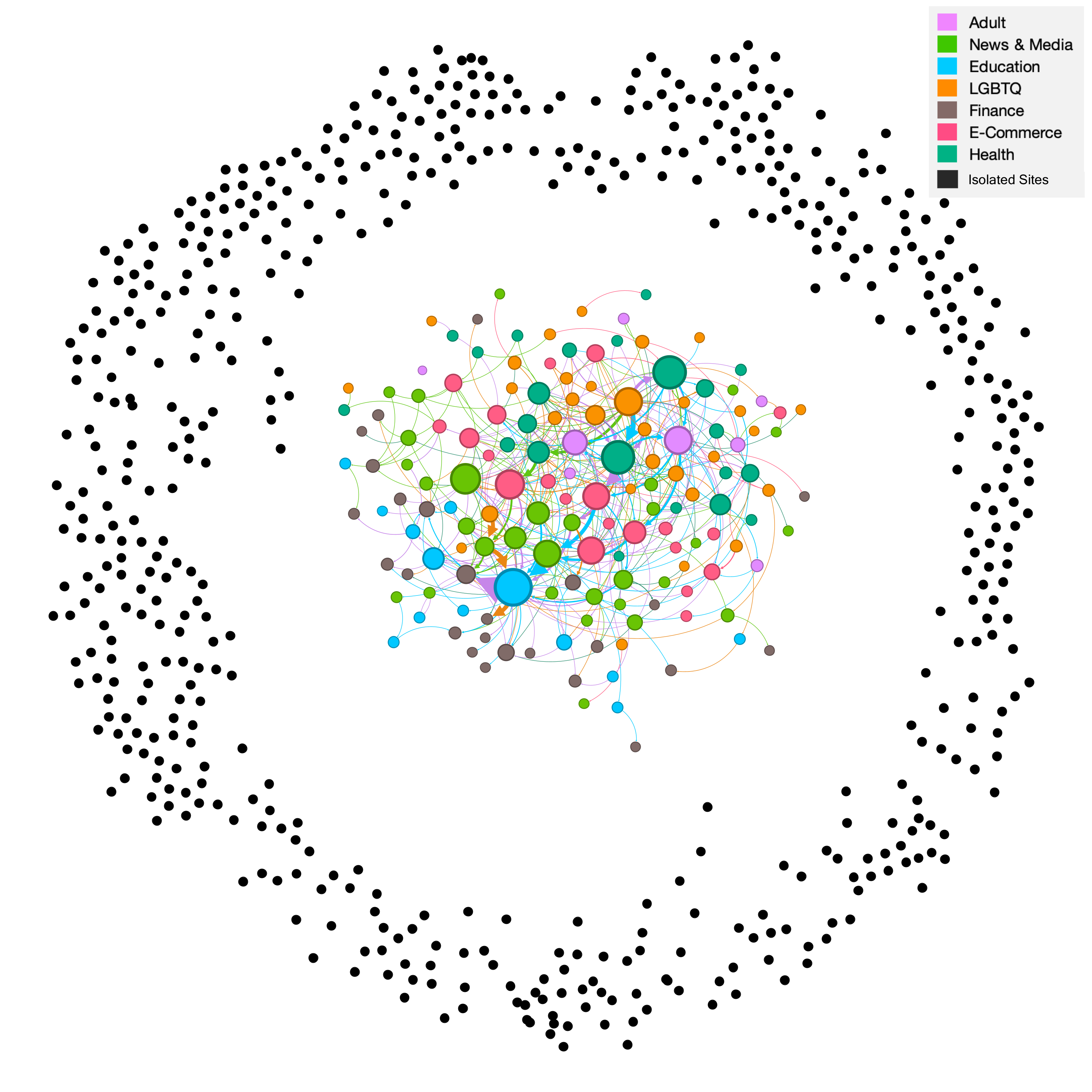}
    \caption{Between-context websites connection, originated from LGBTQ websites. Each vertex is colored based on the calculated chromatic number of the graph.}
    \label{fig:enter-label}
\end{figure}

Overall, to prevent single or multiple context collapse from the LGBTQ context, we need 61 different containers and split first-party websites across those containers based on the lists provided in the appendix. These lists and number of containers should be regularly updated to capture the dynamic nature of Web tracking in order to effectively prevent context collapse for users crawling the top-100 popular LGBTQ websites.

We replicated the analysis for each Web context. To do so, we calculated the average number of colors (=containers) required to prevent persistent identification within each of the studied contexts. The vertex chromatic numbers for within context collapse in each context is the average number across all within context graph networks created in 28 days of data collection for each context. Our results are: \textbf{\textit{Adult}} - 13; \textbf{\textit{eCommerce}} - 40; \textbf{\textit{Education}} - 52; \textbf{\textit{Finance}} - 56; \textbf{\textit{Health}} - 48; \textbf{\textit{LGBTQ}} - 47; and \textbf{\textit{News \& Media}} 67. These are the numbers of containers, on average, required to prevent single context collapse within each context. We will need eight additional containers to prevent multiple context collapse from this context to other contexts in the study. 

Importantly, splitting browser's storage into containers will not prevent persistent identification by stateless identifiers. The JS fingerprinting methods that were studied do not store any data on users' browsers and cannot be blocked by splitting browser's storage to separate containers.

\section{Conclusion}
    
The Web is an essential part of our lives, constructed by an array of contexts, and serves as an enormous source for tracking individuals across their interactions in different contexts. Context collapse on the Web was found to be pervasive both within and across contexts, albeit to different degrees, making it impossible to maintain a fragmented identity without proper protections or meaningful regulation. This is a violation of privacy according to the theory of contextual integrity. Our study aims to empirically assess context collapse by developing a new measurement for Web privacy - the degree of within- and between-context collapse.

A key driving factor of persistent identification patterns are third-party cookie IDs. Even though they get blocked by default across various browsers, Google, holding 65 percent of the browsers' market share as of August 2024 \cite{Statista.2024}, has recently declared that third-party cookies are here to stay \cite{TomsGuide.2024}. We believe our analysis can inform Chrome users, show how another popular persistent identification method - JS fingerprinting - complements cookie IDs, and help users, who are often not privacy tech experts, maintain a fragmented identity on the Web. We have provided a first glimpse on how tracking differs between and within contexts, allowing for a better understanding of the ‘invisible contracts’ we currently have with our websites.

We do not declare that this is an end of our contextual investigation journey of the Web. This is just a first modest step to empirically analyze the behavior of trackers across online contexts based on the theory of privacy as contextual integrity. Looking ahead, we plan to inspect how pervasive cookie-syncing practices and the usage of tracking pixels contribute to context collapse and how different geolocations of browsing and real-user browsing habits differently experience context collapse on the Web.

\bibliographystyle{ACM-Reference-Format}
\bibliography{sample-base}

\appendix

\section{Statistical Tests for Differences Between Contexts}

\subsection{Between-Context Persistent Identifiers}

The difference between the mean numbers of persistent identifiers from each contexts were found to be statistically significant according to the one-sided ANOVA test below. Figure 13 describes all the observations of this number, across contexts, over time, and Figure 14 shows the ANOVA test results.

\begin{figure}[h]
    \centering
    \includegraphics[width=1\linewidth]{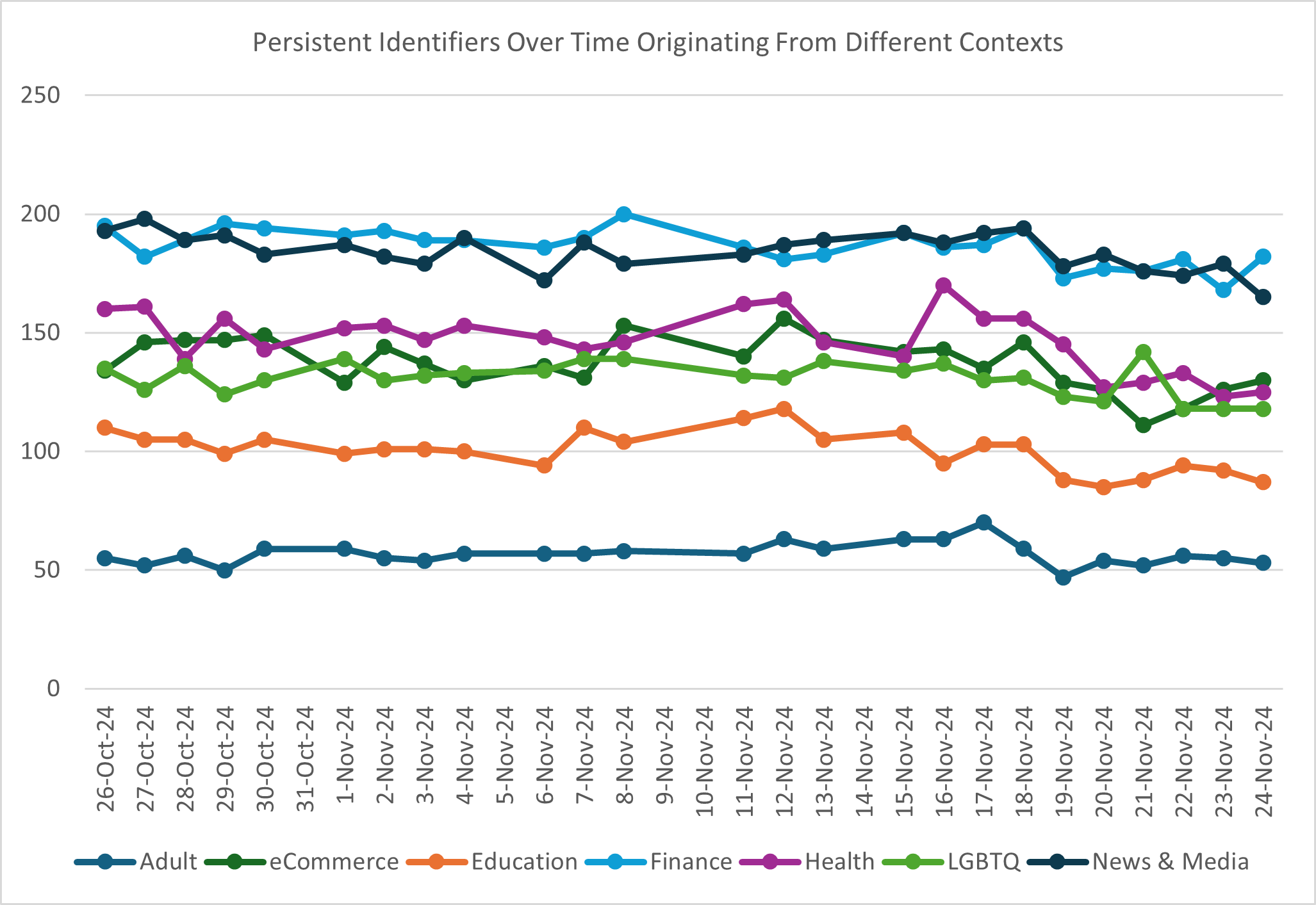}
    \caption{Persistent Identifiers Over Time Originating From Different Contexts}
    \label{fig:enter-label}
\end{figure}

\begin{figure}[h]
    \centering
    \includegraphics[width=1\linewidth]{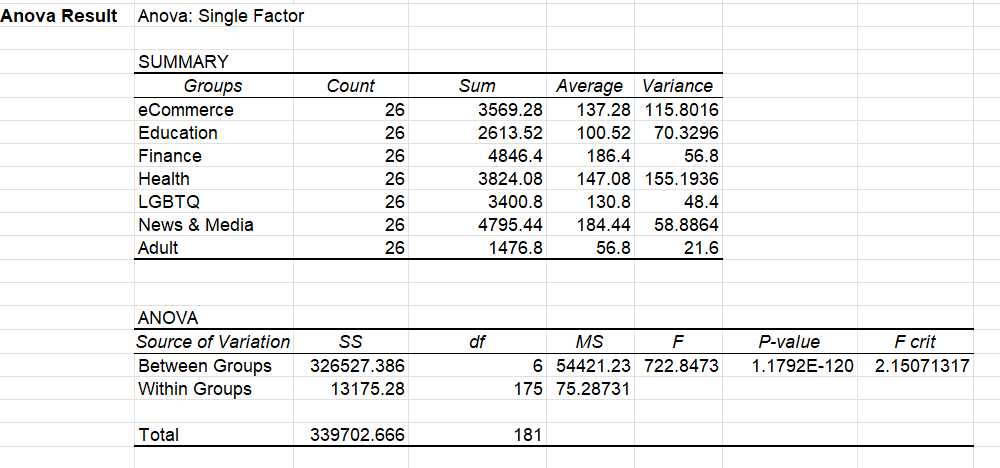}
    \caption{One-sided ANOVA Test Results for Differences Between Means of Persistent Identifiers From Each Context}
    \label{fig:enter-label}
\end{figure}

\subsection{Between-Context Participating Websites}

The difference between the mean numbers of participating websites from each contexts were found to be statistically significant according to the one-sided ANOVA test below. Figure 15 describes all the observations of this number across contexts over time, and Figure 16 shows the ANOVA test results.

\begin{figure}[h]
    \centering
    \includegraphics[width=1\linewidth]{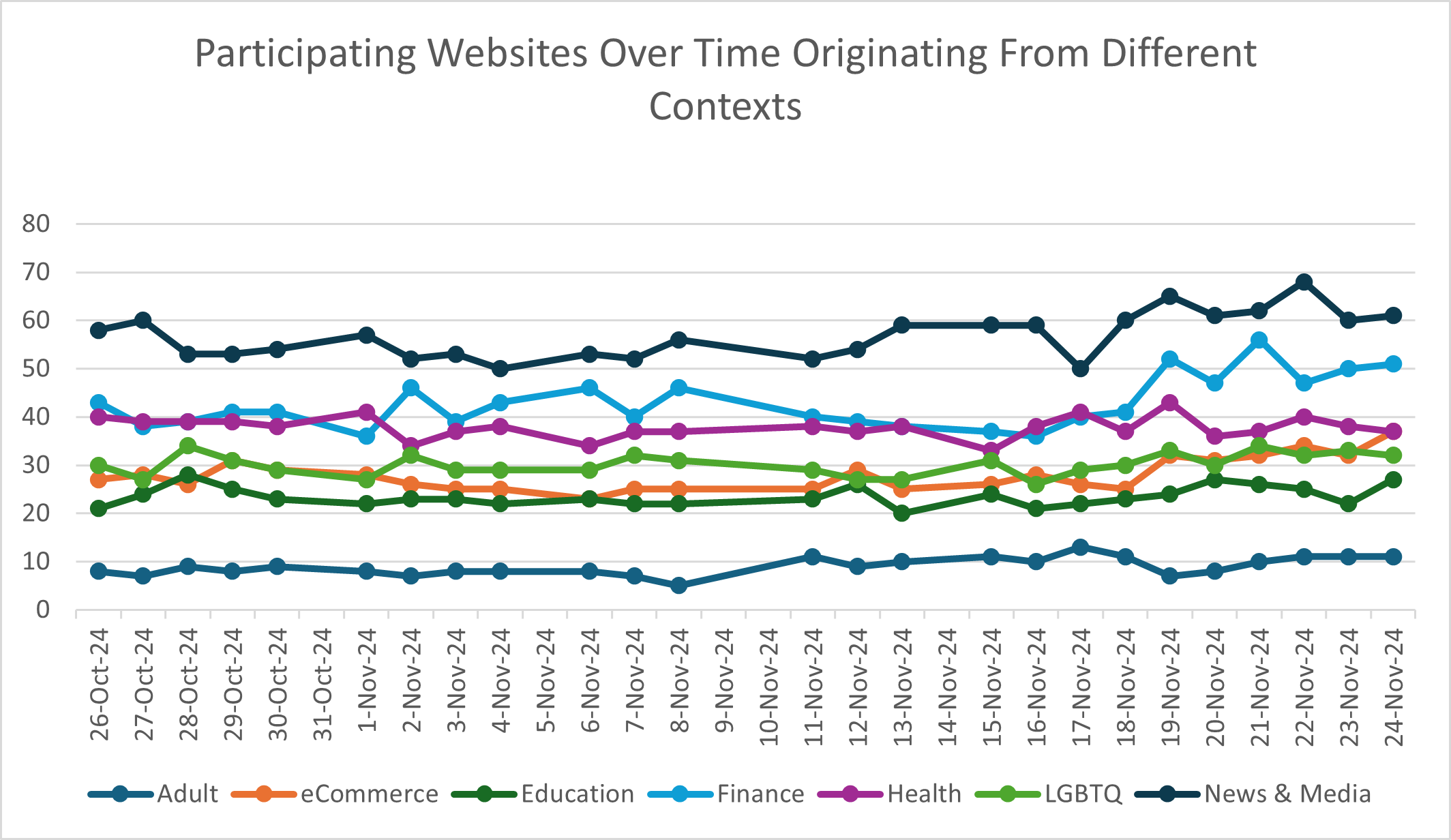}
    \caption{Participating Websites Over Time Originating From Different Contexts}
    \label{fig:enter-label}
\end{figure}

\begin{figure}[h]
    \centering
    \includegraphics[width=1\linewidth]{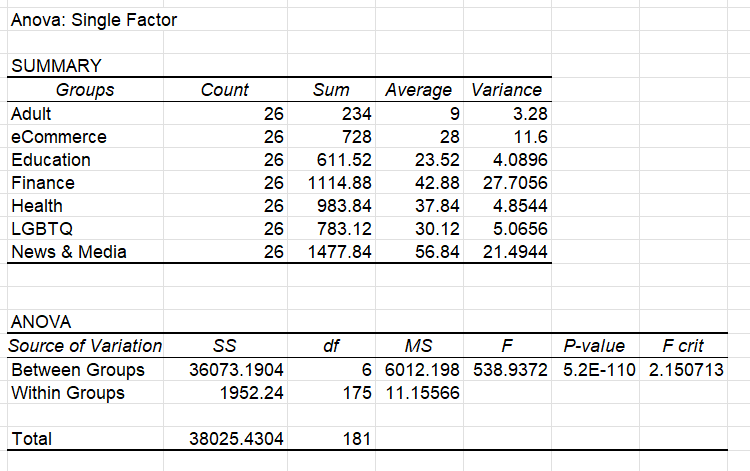}
    \caption{One-sided ANOVA Test Results for Differences Between Means of Participating Websites From Each Context}
    \label{fig:enter-label}
\end{figure}

\subsection{Within-context persistent identifiers}
The difference between the mean numbers of persistent identifiers from each contexts were found to be statistically significant according to the one-sided ANOVA test below. Figure 17 describes all the observations of this value across contexts over time, and Figure 18 shows the ANOVA test results.

\begin{figure}[h]
    \centering
    \includegraphics[width=1\linewidth]{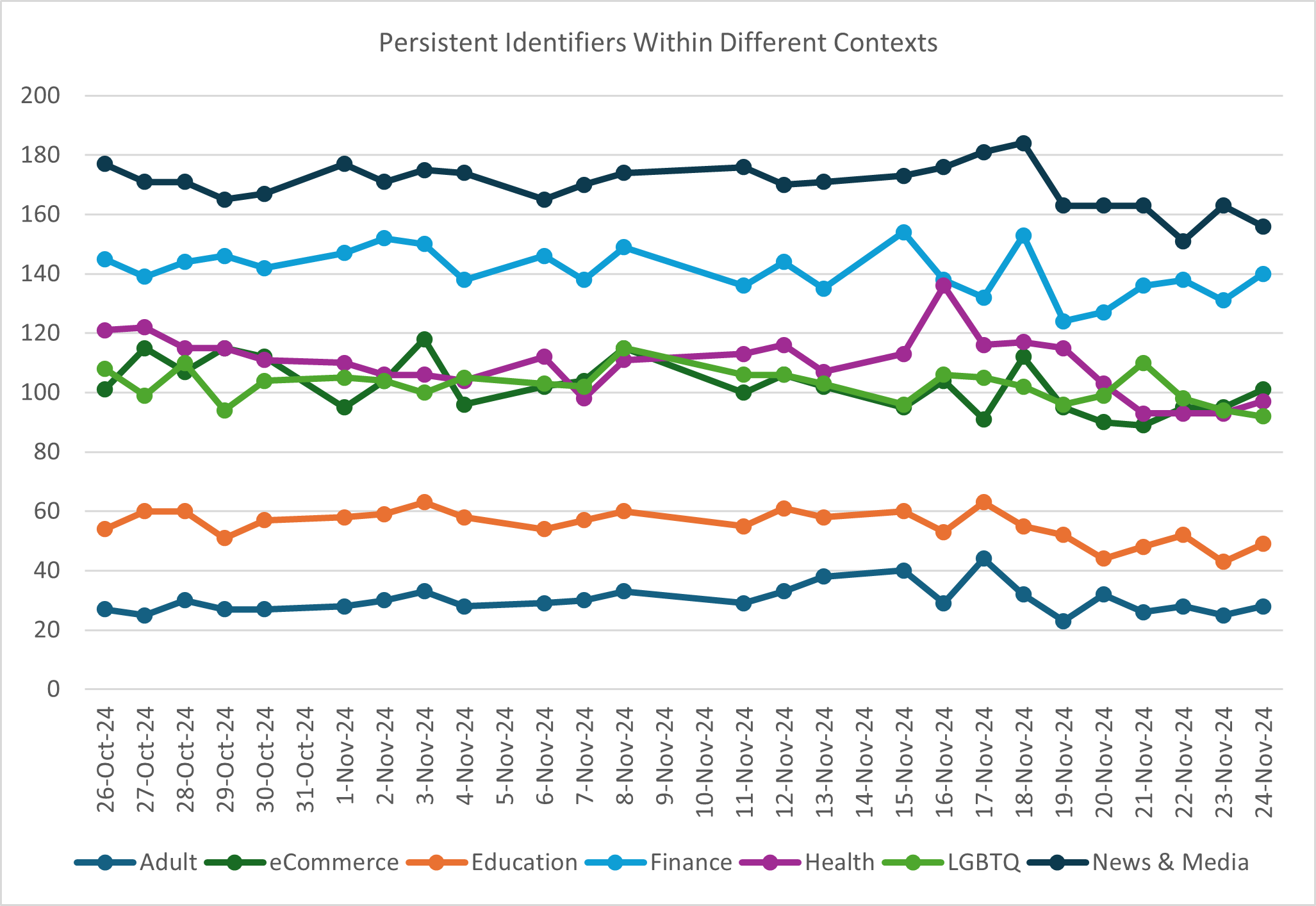}
    \caption{Persistent Identifiers Over Time Originating From Different Contexts for the single context collapse analysis}
    \label{fig:enter-label}
\end{figure}

\begin{figure}[h]
    \centering
    \includegraphics[width=1\linewidth]{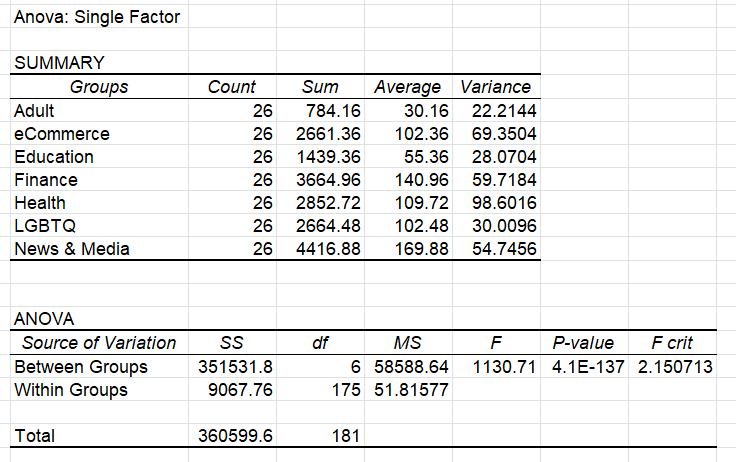}
    \caption{One-sided ANOVA Test Results for Differences Between Means of Persistent Identifiers From Each Context in the single context collapse analysis}
    \label{fig:enter-label}
\end{figure}

\subsection{Within-context participating websites}
The difference between the mean numbers of participating websites from each contexts were found to be statistically significant according to the one-sided ANOVA test below. Figure 19 describes all the observations of this value across contexts over time, and Figure 20 shows the ANOVA test results.

\begin{figure}[h]
    \centering
    \includegraphics[width=1\linewidth]{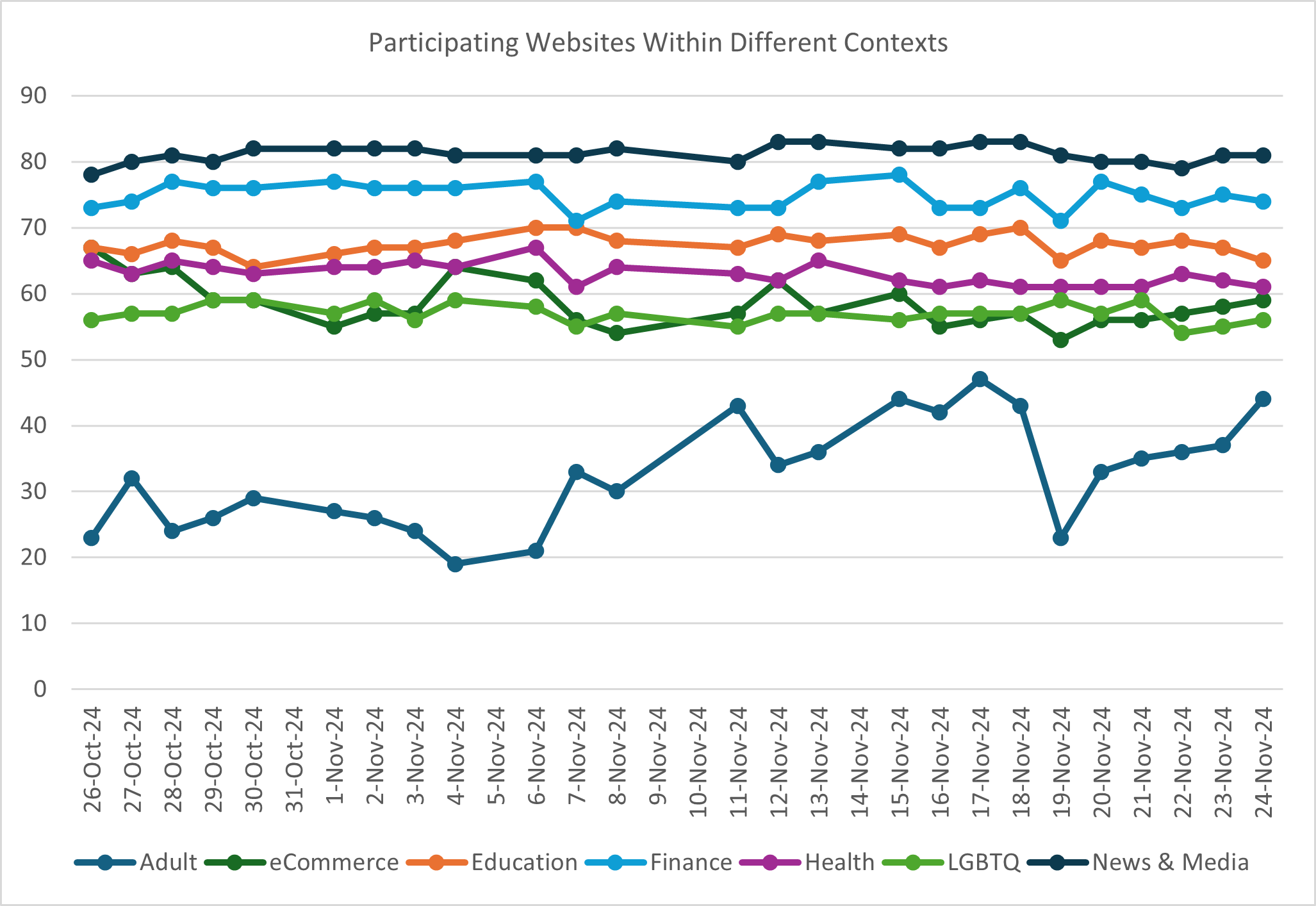}
    \caption{Participating Websites Over Time Originating From Different Contexts for the single context collapse analysis}
    \label{fig:enter-label}
\end{figure}

\begin{figure}[h]
    \centering
    \includegraphics[width=1\linewidth]{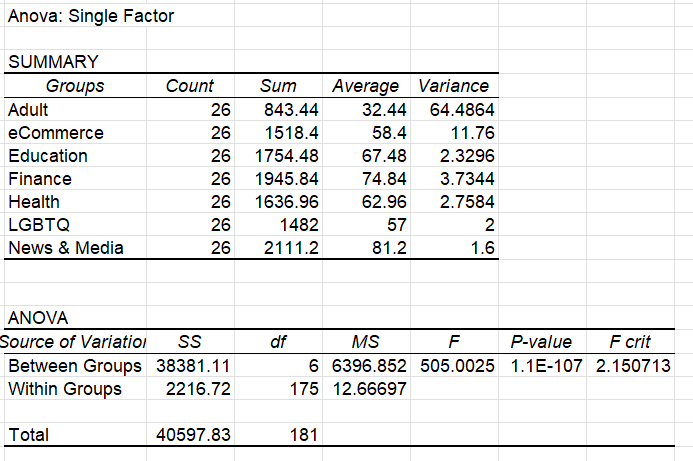}
    \caption{One-sided ANOVA Test Results for Differences Between Means of Participating Websites From Each Context}
    \label{fig:enter-label}
\end{figure}

\section{Appendix Tables}
\begin{table}[h!]
\centering
\scriptsize
\label{tab:styled_table}
\begin{tabular}{|l|l|l|l|l|l|l|}
\hline
\textbf{Health} & \textbf{LGBTQ} & \textbf{E-Commerce} & \textbf{Finance} & \textbf{News \& Media} & \textbf{Adult} & \textbf{Education} \\
\hline
nih.gov & fabguys.com & amazon.com & paypal.com & yahoo.com & pornhub.com & cambridge.org \\
\hline
healthline.com & appnebula.co & temu.com & tradingview.com & yahoo.co.jp & xvideos.com & mit.edu \\
\hline
mayoclinic.org & g4guys.com & ebay.com & chase.com & naver.com & xhamster.com & harvard.edu \\
\hline
clevelandclinic.org & mensnet.jp & amazon.co.jp & capitalone.com & news.yahoo.co.jp & xnxx.com & studentaid.gov \\
\hline
cvs.com & thepinknews.com & aliexpress.com & investing.com & globo.com & rqr.one & iyf.tv \\
\hline
webmd.com & gays-cruising.com & rakuten.co.jp & moneycontrol.com & msn.com & stripchat.com & utoronto.ca \\
\hline
medicalnewstoday.com & bullchat.com & amazon.in & coinmaster.com & cnn.com & chaturbate.com & cuny.edu \\
\hline
walgreens.com & datalounge.com & walmart.com & economictimes.com & bbc.co.uk & eporner.com & joinhandshake.com \\
\hline
1mg.com & queerty.com & amazon.de & citi.com & nytimes.com & spankbang.com & stanford.edu \\
\hline
nhs.uk & advocate.com & etsy.com & bankofamerica.com & qq.com & mnaspm.com & usg.edu \\
\hline
doctolib.fr & joemygod.com & ozon.ru & intuit.com & bbc.com & xnxx.health & ubc.ca \\
\hline
kisskh.co & out.com & amazon.co.uk & americanexpress.com & uol.com.br & faphouse.com & cornell.edu \\
\hline
altibbi.com & autostraddle.com & wildberries.ru & wellsfargo.com & news.google.com & xhamster.desi & shaalaa.com \\
\hline
msdmanuals.com & lgbtqnation.com & avito.ru & hdfcbank.com & theguardian.com & missav.com & purdue.edu \\
\hline
aarp.org & queer.de & flipkart.com & poste.it & foxnews.com & blurbreimbursetrombone.com & elluciancloud.com \\
\hline
medlineplus.gov & hrc.org & coupang.com & fidelity.com & infobae.com & onlyfans.com & berkeley.edu \\
\hline
alodokter.com & pride.com & mercadolivre.com.br & zerodha.com & onet.pl & livejasmin.com & scribbr.com \\
\hline
athenahealth.com & ourtruecolors.org & ebay.co.uk & adp.com & dailymail.co.uk & magsrv.com & umich.edu \\
\hline
tuasaude.com & daleenelarcojuana.com.ar & amazon.it & schwab.com & wp.pl & xvv1deos.com & asu.edu \\
\hline
iherb.com & gay.it & amazon.ca & exness.com & douyin.com & dmm.co.jp & columbia.edu \\
\hline
nhathuoclongchau.com.vn & xl-gaytube.com & amazon.fr & finance.yahoo.co.jp & news18.com & xhamsterlive.com & byu.edu \\
\hline
cdc.gov & gaysir.no & allegro.pl & coinmarketcap.com & news.naver.com & holahupa.com & illinois.edu \\
\hline
halodoc.com & gaybodyblog.com & taobao.com & caixa.gov.br & livedoor.jp & xnxx.es & unsw.edu.au \\
\hline
psychologytoday.com & iboys.cz & shopee.vn & rakuten-sec.co.jp & finance.yahoo.com & txnhh.com & umn.edu \\
\hline
verywellhealth.com & queer.pl & amazon.com.br & trustpilot.com & detik.com & ixxx.com & colorado.edu \\
\hline
abczdrowie.pl & kmhesaplama.com & target.com & sbisec.co.jp & bild.de & cityheaven.net & ufl.edu \\
\hline
drugs.com & xtasis.org & mercari.com & tinkoff.ru & news.yahoo.com & noodlemagazine.com & unisa.ac.za \\
\hline
my-personaltrainer.it & gaycities.com & craigslist.org & klarna.com & people.com & xham.live & panopto.com \\
\hline
doctoralia.com.br & yt5s.is & market.yandex.ru & binance.com & usatoday.com & pornhub.org & utexas.edu \\
\hline
v2i8b.com & easygaychat.com & shopping.yahoo.co.jp & groww.in & auone.jp & youporn.com & upenn.edu \\
\hline
vidal.ru & nakluky.cz & ebay.de & alipay.com & forbes.com & clobberprocurertightwad.com & tamu.edu \\
\hline
goodrx.com & glaad.org & leboncoin.fr & marketwatch.com & 163.com & erome.com & wisc.edu \\
\hline
apteka.ru & adriel.com & rakuten.com & discover.com & indiatimes.com & stripchatgirls.com & estacio.br \\
\hline
prodoctorov.ru & milfswipes.com & shopee.co.id & experian.com & aajtak.in & tnaflix.com & commonapp.org \\
\hline
phreesia.net & gay.blog.br & shopee.com.br & icicibank.com & nypost.com & qorno.com & nyu.edu \\
\hline
health.clevelandclinic.org & erunet.co.jp & kleinanzeigen.de & alfabank.ru & t-online.de & dood.li & anhanguera.com \\
\hline
athena.io & dgdgdg.com & amazon.es & bancoestado.cl & interia.pl & love4porn.com & uba.ar \\
\hline
vinmec.com & gayboy.at & trendyol.com & sberbank.ru & vnexpress.net & xgroovy.com & gptzero.me \\
\hline
sprosivracha.com & qruiser.com & ticketmaster.com & businesstoday.in & india.com & rule34.xxx & mcgill.ca \\
\hline
analdin.xxx & cruisinggays.com & sahibinden.com & wise.com & n-tv.de & 4kporn.xxx & uwaterloo.ca \\
\hline
janeapp.com & lesbicanarias.es & shopee.co.th & synchrony.com & hindustantimes.com & iporntv.net & ucla.edu \\
\hline
medonet.pl & instinctmagazine.com & dmm.com & creditkarma.com & sohu.com & xvideos.es & sydney.edu.au \\
\hline
suckhoedoisong.vn & taimi.com & mercadolibre.com.ar & td.com & livehindustan.com & bongacams.com & rutgers.edu \\
\hline
webteb.com & tgcomics.com & olx.pl & money.pl & rambler.ru & mat6tube.com & fgv.br \\
\hline
menshealth.com & club21.org & costco.com & toyokeizai.net & washingtonpost.com & hetapus.com & psu.edu \\
\hline
tabletki.ua & xtramagazine.com & shopping.naver.com & kontur.ru & elpais.com & redtube.com & wgu.edu \\
\hline
who.int & lesarion.com & shope.ee & etoro.com & corriere.it & pornzog.com & desire2learn.com \\
\hline
womenshealthmag.com & kobiety-kobietom.com & mercadolibre.com.mx & login.gov & repubblica.it & dlsite.com & unimelb.edu.au \\
\hline
imss.gob.mx & transequality.org & amazon.com.mx & rakuten-card.co.jp & rbc.ru & pornhat.one & ratemyprofessors.com \\
\hline
eapteka.ru & equaldex.com & jd.com & syf.com & fmkorea.com & xhamster2.com & hku.hk \\
\hline
babycenter.com & gaydar.net & wayfair.com & nspk.ru & tribunnews.com & xnxx2.com & openstax.org \\
\hline
memorial.com.tr & allout.org & olx.com.br & vtb.ru & newsweek.com & sxyprn.com & yale.edu \\
\hline
hopkinsmedicine.org & erininthemorning.com & kakaku.com & eastmoney.com & cnbc.com & nhentai.net & uh.edu \\
\hline
invitro.ru & washingtonblade.com & slickdeals.net & foxbusiness.com & elmundo.es & sex-studentki.link & ignou.ac.in \\
\hline
rlsnet.ru & qx.se & aliexpress.ru & credit-agricole.fr & nbcnews.com & crazyporn.xxx & liberty.edu \\
\hline
shop-apotheke.com & dnamagazine.com.au & alibaba.com & mufg.jp & independent.co.uk & crazyvideotodownload.com & unicesumar.edu.br \\
\hline
drogaraia.com.br & tslove.net & hepsiburada.com & investopedia.com & livedoor.com & smkezc.com & suny.edu \\
\hline
ecwcloud.com & pinknews.co.uk & tokopedia.com & moneyforward.com & apnews.com & youjizz.com & ucf.edu \\
\hline
chemistwarehouse.com.au & bakala.org & bol.com & mercadopago.com.ar & goo.ne.jp & brazzersnetwork.com & uq.edu.au \\
\hline
simplepractice.com & cromosomax.com & dns-shop.ru & ft.com & indianexpress.com & pornpics.com & libgen.li \\
\hline
questdiagnostics.com & b-gay.com & tmall.com & fortune.com & businessinsider.com & twinrdsyte.com & ucsd.edu \\
\hline
practo.com & primarius.app & amazon.com.au & ideal.nl & ndtv.com & hdtube.porn & osu.edu \\
\hline
wlgrn.com & stonewall.org.uk & magazineluiza.com.br & adyen.com & ukr.net & hot-sex-tube.com & utah.edu \\
\hline
medscape.com & portugalgay.pt & indiamart.com & cafef.vn & cbsnews.com & theporndude.com & monash.edu \\
\hline
apollopharmacy.in & gayua.com & olx.ua & bajajfinserv.in & lenta.ru & xnxxcom.xyz & snhu.edu \\
\hline
drogasil.com.br & sissykiss.com & shopee.tw & xe.com & eenadu.net & hqporner.com & ucdavis.edu \\
\hline
health.com & bi.org & shopee.com.my & goodreturns.in & lefigaro.fr & cambaddies.cc & moe.edu.tw \\
\hline
labcorp.com & gaychat.nl & shopee.ph & sberbank.com & abplive.com & xxxvideo.link & usc.edu \\
\hline
aamc.org & queermajority.com & subito.it & ilsole24ore.com & oneindia.com & hitomi.la & ox.ac.uk \\
\hline
uptodate.com & fridae.asia & my-best.com & progressive.com & lanacion.com.ar & sexvid.pro & duke.edu \\
\hline
parenting.pl & thenewcivilrightsmovement.com & otto.de & mitid.dk & novinky.cz & jable.tv & virginia.edu \\
\hline
uworld.com & lifeout.com & eventbrite.com & coinbase.com & francetvinfo.fr & fuq.com & ncsu.edu \\
\hline
doctolib.de & gay-szene.net & kohls.com & banki.ru & o2.pl & erothots1.com & msu.edu \\
\hline
health.mail.ru & gayromeo.com & amzn.to & check24.de & news.mail.ru & porno365.scot & unc.edu \\
\hline
medlatec.vn & rainbownet.jp & amazon.sa & nguoiquansat.vn & terra.com.br & fapello.com & gmu.edu \\
\hline
naxadrug.com & rainbowrailroad.org & alibaba-inc.com & nerdwallet.com & indiatoday.in & eroterest.net & niche.com \\
\hline
rxinform.org & ciciful.com & poshmark.com & screener.in & ria.ru & xnxx.tv & pak-mcqs.net \\
\hline
babyblog.ru & domsubliving.com & idealo.de & forexfactory.com & nikkei.com & 2526june2024.com & nus.edu.sg \\
\hline
athomedaily.com & manhunt.com & rozetka.com.ua & short.gy & kompas.com & thisvid.com & jhu.edu \\
\hline
consultaremedios.com.br & ilga.org & gmarket.co.kr & kaiserpermanente.org & hurriyet.com.tr & fetlife.com & admissions.nic.in \\
\hline
uteka.ru & towleroad.com & otomoto.pl & usbank.com & reuters.com & doorblog.jp & kuleuven.be \\
\hline
znanylekarz.pl & coat.co.jp & argos.co.uk & nikkei225jp.com & vg.no & getallmylinks.com & iu.edu \\
\hline
gemotest.ru & byren.cn & samsclub.com & robinhood.com & lemonde.fr & a64x.com & anthropic.com \\
\hline
verywellmind.com & locuragay.com & marktplaats.nl & whitebit.com & tagesschau.de & tktube.com & bu.edu \\
\hline
myfitnesspal.com & crossdresserheaven.com & ebay.it & bb.com.br & nu.nl & 2520june2024.com & gcu.edu \\
\hline
\end{tabular}
\end{table}

\clearpage

\begin{table}[h!]
\centering
\scriptsize
\label{tab:styled_table}
\begin{tabular}{|l|l|l|l|l|l|l|}
\hline
\textbf{Health} & \textbf{LGBTQ} & \textbf{E-Commerce} & \textbf{Finance} & \textbf{News \& Media} & \textbf{Adult} & \textbf{Education} \\
\hline
racunn.com & hayunalesbianaenmisopa.com & capitaloneshopping.com & business-standard.com & buzzfeed.com & txxx.com & ualberta.ca \\
\hline
fda.gov & jackd.com & shp.ee & pse.com.co & kp.ru & boyfriendtv.com & umd.edu \\
\hline
healthgrades.com & genxy-net.com & 11st.co.kr & razorpay.com & substack.com & cam4.com & yorku.ca \\
\hline
whattoexpect.com & burnettfoundation.org.nz & megamarket.ru & serasa.com.br & clarin.com & xnxx-arabic.com & uchicago.edu \\
\hline
hesaplama.net & thaiboyslove.com & ssg.com & royalbank.com & wsj.com & xvideos2.com & libgen.rs \\
\hline
doctoralia.com.mx & l-mag.de & akakce.com & citibankonline.com & iltalehti.fi & 4chan.org & utn.edu.ar \\
\hline
ubie.app & daddyhunt.com & 1688.com & dpbossss.services & welt.de & rtbbtech.com & du.ac.in \\
\hline
gsuplementos.com.br & deadwildroses.com & lazada.co.th & altin.in & yomiuri.co.jp & simpcity.su & pitt.edu \\
\hline
lww.com & awrymenswear.com & ebay.com.au & smbc-card.com & ouest-france.fr & aznude.com & washington.edu \\
\hline
krasotaimedicina.ru & transgendermap.com & amazon.com.tr & moneysavingexpert.com & dailyhunt.in & supjav.com & tophat.com \\
\hline
pharmeasy.in & heckinunicorn.com & mercadolibre.com.co & pnc.com & ynet.co.il & best-video-app.com & sena.edu.co \\
\hline
mp.pl & susans.org & markt.de & cnbctv18.com & sozcu.com.tr & xv-ru.com & vt.edu \\
\hline
doz.pl & games.mashable.com & meesho.com & ing.nl & spiegel.de & movie.eroterest.net & cuhk.edu.hk \\
\hline
lybrate.com & emptyclosets.com & valuecommerce.com & benzinga.com & edition.cnn.com & xnxx.com.se & rochester.edu \\
\hline
doxy.me & readyforpolyamory.com & stubhub.com & finviz.com & amarujala.com & dinotube.com & thestudentro \\
\hline
\end{tabular}
\caption{Top Popular Websites Used for Data Collection}
\end{table}

\begin{table*}[h]
\centering
\scriptsize
\begin{tabular}{|>{\centering\arraybackslash}p{0.1\linewidth}|>{\raggedright\arraybackslash}p{0.8\linewidth}|}
\hline
\textbf{Color} & \textbf{First-party websites} \\ \hline
0 & [daddyhunt.com, fridae.asia] \\ \hline
1 & [cruisinggays.com] \\ \hline
2 & [gay.it] \\ \hline
3 & [instinctmagazine.com] \\ \hline
4 & [thenewcivilrightsmovement.com] \\ \hline
5 & [joemygod.com] \\ \hline
6 & [qruiser.com] \\ \hline
7 & [b-gay.com] \\ \hline
8 & [rainbownet.jp, transgendermap.com] \\ \hline
9 & [gaychat.nl] \\ \hline
10 & [transequality.org, erunet.co.jp] \\ \hline
11 & [rainbowrailroad.org] \\ \hline
12 & [deadwildroses.com] \\ \hline
13 & [queerty.com] \\ \hline
14 & [genxy-net.com] \\ \hline
15 & [hayunalesbianaenmisopa.com] \\ \hline
16 & [portugalgay.pt] \\ \hline
17 & [advocate.com] \\ \hline
18 & [hrc.org] \\ \hline
19 & [xtramagazine.com] \\ \hline
20 & [nakluky.cz] \\ \hline
21 & [cromosomax.com] \\ \hline
22 & [thepinknews.com] \\ \hline
23 & [pinknews.co.uk] \\ \hline
24 & [lesbicanarias.es] \\ \hline
25 & [l-mag.de] \\ \hline
26 & [awrymenswear.com] \\ \hline
27 & [thaiboyslove.com] \\ \hline
28 & [ilga.org] \\ \hline
29 & [iboys.cz] \\ \hline
30 & [queer.pl] \\ \hline
31 & [kmhesaplama.com] \\ \hline
32 & [domsubliving.com] \\ \hline
33 & [autostraddle.com] \\ \hline
34 & [byren.cn] \\ \hline
35 & [pride.com] \\ \hline
36 & [burnettfoundation.org.nz] \\ \hline
37 & [washingtonblade.com] \\ \hline
38 & [gaydar.net] \\ \hline
39 & [dnamagazine.com.au] \\ \hline
40 & [out.com] \\ \hline
41 & [bi.org] \\ \hline
42 & [adriel.com, gay.blog.br] \\ \hline
43 & [heckinunicorn.com] \\ \hline
44 & [datalounge.com] \\ \hline
45 & [gaycities.com] \\ \hline
46 & [allout.org, stonewall.org.uk] \\ \hline
47 & [manhunt.com] \\ \hline
48 & [erininthemorning.com] \\ \hline
49 & [glaad.org] \\ \hline
50 & [towleroad.com] \\ \hline
51 & [lgbtqnation.com] \\ \hline
52 & [ciciful.com] \\ \hline
53 & [fabguys.com, appnebula.co, g4guys.com, mensnet.jp, gays-cruising.com, bullchat.com, queer.de, ourtruecolors.org, daleenelarcojuana.com.ar, xl-gaytube.com, gaysir.no, gaybodyblog.com, xtasis.org, yt5s.is, easygaychat.com, milfswipes.com, dgdgdg.com, gayboy.at, taimi.com, tgcomics.com, club21.org, lesarion.com, kobiety-kobietom.com, equaldex.com, qx.se, tslove.net, bakala.org, primarius.app, gayua.com, sissykiss.com, queermajority.com, gay-szene.net, gayromeo.com, coat.co.jp, locuragay.com, crossdresserheaven.com, jackd.com, susans.org, games.mashable.com, emptyclosets.com, readyforpolyamory.com] \\ \hline
\end{tabular}
\caption{Container assignment based on graph coloring scheme to prevent single context collapse within top-100 popular LGBTQ websites}
\label{tab:grouped_websites}
\end{table*} 

\begin{table*}[h]
\centering
\begin{tabular}{|>{\centering\arraybackslash}p{0.1\linewidth}|>{\raggedright\arraybackslash}p{0.8\linewidth}|}
\hline
\textbf{Color} & \textbf{First-party websites} \\ \hline
1 & [ciciful.com, lgbtqnation.com, towleroad.com, glaad.org, allout.org, gaycities.com, adriel.com, out.com, gaydar.net, washingtonblade.com, pride.com, autostraddle.com, iboys.cz, thaiboyslove.com, l-mag.de, lesbicanarias.es, cromosomax.com, nakluky.cz, xtramagazine.com, hrc.org, advocate.com, genxy-net.com, queerty.com, rainbownet.jp, b-gay.com, joemygod.com, thenewcivilrightsmovement.com, instinctmagazine.com, cruisinggays.com, fridae.asia] \\ \hline
2 & [aarp.org, 1mg.com, consultaremedios.com.br, memorial.com.tr, whattoexpect.com, verywellhealth.com, suckhoedoisong.vn, athomedaily.com, medscape.com, athenahealth.com, apollopharmacy.in, gsuplementos.com.br, pharmeasy.in, womenshealthmag.com, health.com, babycenter.com, drugs.com, tuasaude.com, invitro.ru, altibbi.com, chemistwarehouse.com.au, hesaplama.net] \\ \hline
3 & [dmm.co.jp, tktube.com, livejasmin.com, doorblog.jp, iporntv.net, dlsite.com, erothots1.com] \\ \hline
4 & [amazon.fr, flipkart.com, target.com, mercari.com, walmart.com, tokopedia.com, gmarket.co.kr, amazon.in, ssg.com, marktplaats.nl, 11st.co.kr, kakaku.com, dmm.com, rakuten.co.jp, olx.com.br, slickdeals.net, rakuten.com, wayfair.com, aliexpress.ru, kohls.com, rozetka.com.ua] \\ \hline
5 & [medonet.pl, elpais.com, o2.pl, forbes.com, auone.jp, ukr.net, eenadu.net, apnews.com, theguardian.com, yomiuri.co.jp, vnexpress.net, lanacion.com.ar, usatoday.com, businessinsider.com, foxnews.com, ndtv.com, cnn.com, buzzfeed.com, nbcnews.com, www.edition.cnn.com, india.com, detik.com, cnbc.com, cbsnews.com, ria.ru, ynet.co.il, kp.ru, independent.co.uk, msn.com, hurriyet.com.tr] \\ \hline
6 & [uh.edu, ucdavis.edu, mcgill.ca, ratemyprofessors.com, asu.edu, unicesumar.edu.br, snhu.edu, jhu.edu, wgu.edu, estacio.br, joinhandshake.com, nyu.edu, rutgers.edu] \\ \hline
7 & [serasa.com.br, discover.com, intuit.com, benzinga.com, bajajfinserv.in, nguoiquansat.vn, investopedia.com, xe.com, moneyforward.com, banki.ru, foxbusiness.com, experian.com, moneycontrol.com, smbc-card.com, marketwatch.com, goodreturns.in, sbisec.co.jp, finviz.com, fortune.com, economictimes.com, businesstoday.in, business-standard.com, mercadopago.com.ar, ilsole24ore.com, adyen.com] \\ \hline
\end{tabular}
\caption{Container assignment based on graph coloring scheme to prevent multiple contexts collapse from the top-100 popular LGBTQ websites}
\label{tab:grouped_websites}
\end{table*}
\begin{table*}[h]
    \centering
    \label{tab:host_info}
    \begin{tabular}{@{}>{\raggedright\arraybackslash}p{0.2\linewidth}>{\raggedright\arraybackslash}p{0.2\linewidth}>{\raggedright\arraybackslash}p{0.3\linewidth}>{\raggedright\arraybackslash}p{0.2\linewidth}@{}}
    \toprule
    \textbf{Website that is also a Persistent Identifier} & \textbf{Category of Website} & \textbf{Categories Found Tracking on} & \textbf{Tracking Instrument: Cookie ID, Javascript Fingerprinting, or Both} \\ 
    \midrule
    market.yandex.ru & E-Commerce & \{Adult, E-Commerce, Finance, LGBTQ, Education, News \& Media, Health\} & Cookie \\ \hline
    news.google.com & News \& Media & \{Adult, E-Commerce, Finance, LGBTQ, Education, News \& Media, Health\} & Both \\ \hline
    magsrv.com & Adult & \{Adult, LGBTQ\} & Cookie \\ \hline
    mail.ru & \{Health, News \& Media\} & \{Adult, E-Commerce, Finance, LGBTQ, News \& Media, Health\} & Both \\ \hline
    paypal.com & Finance & \{E-Commerce, Health\} & Cookie \\ \hline
    mercadolibre.com & E-Commerce & \{E-Commerce, Finance\} & Cookie \\ 
    \bottomrule
    \end{tabular}
    \caption{First-party Websites that also act as Persistent Identifiers}
\end{table*}

\end{document}